\documentclass[journal]{IEEEtran}
\usepackage[T1]{fontenc}% optional T1 font encoding
\usepackage{amsmath,amsfonts}
\usepackage{algorithm}
\usepackage{algpseudocode}
\usepackage{array}
\usepackage[caption=false]{subfig}
\usepackage{textcomp}
\usepackage{stfloats}
\usepackage{url}
\usepackage{verbatim}
\usepackage{graphicx}
\usepackage{cite}
\usepackage[cmintegrals]{newtxmath}
\usepackage{graphicx}
\usepackage{xcolor} 
\usepackage{epsf}
\usepackage{latexsym}
\usepackage{multirow}
\usepackage{multicol}
\usepackage{float}
\usepackage{lipsum}
\usepackage{cleveref}
\usepackage{stackengine}

\newtheorem{lemma}{Lemma}
\newtheorem{assumption}{Assumption}

\newtheorem{theorem}{Theorem}
\newtheorem{proposition}{Proposition}

\newcommand{\taucomm}{\tau^{\text{comm}}}
\newcommand{\taucomp}{\tau^{\text{comp}}}

\newcommand{\argmin}{\operatornamewithlimits{argmin}}
\def\delequal{\mathrel{\ensurestackMath{\stackon[1pt]{=}{\scriptstyle\Delta}}}}

\renewcommand{\algorithmicrequire}{\textbf{Input:}}

\ifCLASSINFOpdf
\else
\fi
\usepackage{amsmath}
\usepackage[cmintegrals]{newtxmath}
\begin{document}
	
	\title{Asynchronous Federated Learning Using Outdated Local Updates Over TDMA Channel}
	
	\author{Jaeyoung~Song, ~\IEEEmembership{Member,~IEEE} and Jun-Pyo Hong, ~\IEEEmembership{Member,~IEEE}% <-this % stops a space
		\thanks{J. Song is with the Department of Electronics Engineering, Pusan National University, Pusan, 46241, Korea (e-mail : jsong@pnu.edu)} \\
		\thanks{Jun-Pyo Hong is with the School of Electronics and Electrical Engineering, Hongik University, Seoul, 04066, Korea (e-mail: jp\_hong@hongik.ac.kr)} }
	
	%\markboth{Journal of \LaTeX\ Class Files,~Vol.~14, No.~8, August~2015}%
	%{Shell \MakeLowercase{\textit{et al.}}: Bare Demo of IEEEtran.cls for IEEE Communications Society Journals}
	
	\maketitle
	
	% As a general rule, do not put math, special symbols or citations
	% in the abstract or keywords.
	\begin{abstract}
		In this paper, we consider asynchronous federated learning (FL) over time-division multiple access (TDMA)-based communication networks.
		Considering TDMA for transmitting local updates can introduce significant delays to conventional synchronous FL, where all devices start local training from a common global model.	In the proposed asynchronous FL approach, we partition devices into multiple TDMA groups, enabling simultaneous local computation and communication across different groups. This enhances time efficiency at the expense of staleness of local updates. We derive the relationship between the staleness of local updates and the size of the TDMA group in a training round. Moreover, our convergence analysis shows that although outdated local updates hinder appropriate global model updates, asynchronous FL over the TDMA channel converges even in the presence of data heterogeneity. Notably, the analysis identifies the impact of outdated local updates on convergence rate.
		Based on observations from our convergence rate, we refine asynchronous FL strategy by introducing an intentional delay in local training. 
		This refinement accelerates the convergence by reducing the staleness of local updates.
		Our extensive simulation results demonstrate that asynchronous FL with the intentional delay can rapidly reduce global loss by lowering the staleness of local updates in resource-limited wireless communication networks.
	\end{abstract}
	
	% Note that keywords are not normally used for peerreview papers.
	
	\IEEEpeerreviewmaketitle
	\section{Introduction}
	The rapid proliferation of edge devices, such as Internet of Things (IoT) devices, has led to the generation of vast amounts of data. As the 
	performance of machine learning models, particularly those involving artificial intelligence (AI), is highly dependent on the size of the datasets used 
	during the training phase, the enormous datasets generated by these edge devices are invaluable to organizations operating AI-driven applications.
	
	However, data generated by edge devices often contains sensitive personal information, such as health or financial data, which raises significant 
	privacy concerns. Additionally, the collection and centralized processing of such large datasets can incur substantial costs. To address these 
	challenges, federated learning (FL) has emerged as a promising solution among researchers \cite{pmlr-v54-mcmahan17a}. 
	
	An FL system typically consists of a server and multiple edge devices. In FL, direct access to the distributed local data by the server or other devices 
	is restricted. Instead, each device updates its model locally using its own data. These local updates are then sent to the server, where they are 
	aggregated to perform a global update. The updated global model is subsequently sent back to the devices for next round.
	Because FL can improve a model without requiring centralized access to all data, communication between edge devices and the server is crucial. Consequently, communication becomes a dominant process in the overall procedure of the FL system, often serving as the bottleneck \cite{konevcny2016federated}.
	
	To address this bottleneck, several studies have explored communication-efficient FL strategies. One such approach, introduced in \cite{stich2018local, yu2019parallel}, employs local stochastic gradient descent (SGD), where multiple updates are performed on the devices before reporting them to the server, thereby reducing communication frequency. Other approaches, such as using quantization techniques to reduce the size of communication 
	payloads, have been proposed in \cite{reisizadeh2020fedpaq}. In \cite{pmlr-v54-mcmahan17a}, federated learning strategy which selects a subset of clients is proposed to reduce communication cost. To further accelerate convergence, 
	researchers in \cite{shi2022talk, song2024optimal}  explored dynamically adjusting batch sizes during FL. 
	
	When FL is conducted over wireless channels, a proper allocation of communication resource can also improve communication efficiency of FL. Several studies have discussed energy consumption \cite{chen2020joint}, number of devices \cite{song2020wireless}, device scheduling \cite{yang2019scheduling}, bandwidth and power allocation \cite{ji2023joint, hong2023ICC, hong2023TWC} for resource-efficient FL over wireless networks. In particular, training time is heavily dependent on the multiple access scheme used for local update aggregation. Various multiple access schemes and their impact on training time were studied, including frequency-division multiple access (FDMA) \cite{shi2022talk, chen2020joint}, time-division multiple access (TDMA) \cite{xu2024latency, hu2020device, bouzinis2023wireless}, and non-orthogonal multiple access (NOMA) \cite{da2022multichannel, sun2020adaptive}.
	
	However, in the aforementioned works, all devices are required to be synchronized; which starts local training using the shared global model at the end of previous training round. Moreover, devices communicates when all devices have completed local training. Thereby, in each training round, computation and communications are strictly divided. When devices have different computation load or capability, devices which finish local computation early need to wait for other devices often called as stragglers to keep synchronicity.
	
	To resolve straggler problem, some studies have proposed asynchronous FL \cite{lian2018asynchronous, 
		koloskova2022sharper, wang2022asynchronous, hu2023scheduling}. 
	In \cite{lian2018asynchronous}, asynchronous SGD is shown to achieve the same convergence rate of synchronous SGD given that the maximum asynchronicity is bounded. 
	The effect of asynchronicity on the convergence rate was discussed in \cite{koloskova2022sharper}. 
	When devices can select to use the most recent global model or global model received in the past, a decision policy was proposed in 
	\cite{wang2022asynchronous}. When global model is periodically aggregated from devices which complete their local 
	updates only, scheduling policy dependent on wireless channel and training data distribution was proposed in 
	\cite{hu2023scheduling}.
	
	However, the impact of the communication protocol on asynchronous FL has not yet been figured out. 
	In fact, in asynchronous FL, 
	the convergence speed depends on the level of asynchronicity between models that devices and the server have. 
	Moreover, the asynchronicity is determined by the communication frequency of each device, which is affected by communication scheme critically. 
	
	To address this, in this paper, we study the impact of TDMA on asynchronous FL where large number of IoT devices hold heterogeneous data. 
	To the best of our knowledge, our work is the first to consider asynchronous FL over TDMA channel. In asynchronous FL, global model can be updated while some of device are performing local training. Thus, there is a difference between the shared global model used for local training and the global model to which the local update is applied. In other words, somewhat outdated local updates are used to update the global model in our asynchronous FL. The staleness of local updates depends on the number of global updates occurred after the device has received global model from the server most recently. When TDMA is used for asynchronous FL, we first demonstrate that the size of the communication group directly affects the timeliness of local updates.
	Furthermore, we analyze the convergence rate of asynchronous FL with outdated local updates, deriving how outdated local updates negatively impact convergence.
	To mitigate this, we propose a strategy that introduces an intentional delay for receiving the global model to minimize staleness without additional cost.
	
	In short, our contribution can be summarized as four-fold.
	\begin{itemize}
		\item We investigate asynchronous FL, which allows devices to perform local training while others communicate, enabling gains from parallel processing of communication and computation, as well as the benefits of distributed computing across multiple devices.
		\item When asynchronous FL is performed using stochastic gradient with constant delay under heterogeneous data distribution, we derive the closed-form expression of convergence rate in terms of the delay of stochastic gradient.
		\item To mitigate staleness of local updates, we propose a strategy which imposes an intentional delay on global model reception, ensuring local training starts from the latest global model as possible. We show that this approach effectively reduces update errors from outdated local updates and accelerates convergence.
		\item Through extensive experiments on real-world datasets, we demonstrate the impact of delay and the performance gains of the proposed strategy across different TDMA group sizes.
	\end{itemize}
	The rest of this paper is organized as follows. 
	In Section \ref{sec:system_model}, we describe the system model considered in this paper. 
	The details of asynchronous FL over TDMA channel are presented in Section \ref{sec:asynch}. 
	Section \ref{sec:convergence} provides convergence analysis of asynchronous FL with outdated local updates. 
	Based on the convergence analysis, we propose the asynchronous FL using intentional delay in Section 
	\ref{sec:int_delay}.  
	Moreover, we present results of experiment using real dataset in Section \ref{sec:exp}. Finally, Section \ref{sec:conclusion} concludes 
	the paper.
	
	\section{System Model}\label{sec:system_model}
	\begin{figure}[t!]
		\centering
		\includegraphics[scale=0.5]{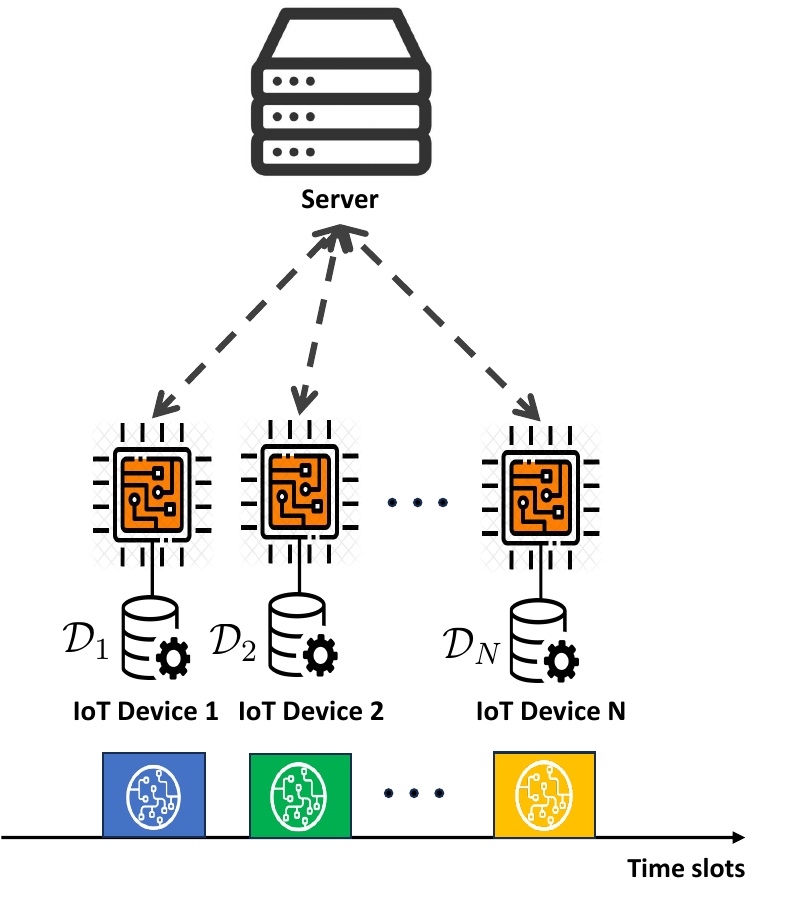}
		\caption{Federated learning system with $N$ IoT devices and a server}\label{fig:system_model}
	\end{figure}
	
	We consider FL over a wireless network which consists of a server and $N$ IoT devices indexed by $n \in \mathcal{N} = \left\lbrace 1,2,\dots, N \right\rbrace$ as shown in the Fig \ref{fig:system_model}. 
	The objective of the FL is to obtain a global model parameterized 
	by a $m$-dimensional real-valued vector $\mathbf{w}$  which fits to distributive dataset across $N$ IoT devices within finite time. 
	Each IoT device $n$ holds private local data $\mathcal{D}_n$ with size $D_n = |\mathcal{D}_n|$.
	The entire dataset stored in $N$ IoT devices is represented as $\mathcal{D} = 
	\cup_{n=1}^N \mathcal{D}_n$. 
	Similarly, we represent $D = \sum_{n=1}^N D_n$ as total number of data samples used in FL.  
	Given that a global model is parameterized by a vector $\mathbf{w}$ , global loss function is defined as
	\begin{align}
		f(\mathbf{w} ) = \frac{1}{D} \sum_{\zeta \in \mathcal{D}} l (\mathbf{w}, \zeta), 
	\end{align} where $\zeta$ is a data sample and $l(\mathbf{w}, \zeta)$ is a loss function defined for each data sample. 
	Moreover, the local loss function of IoT device $n$ can be written as 
	\begin{align}
		f_n(\mathbf{w} ) = \frac{1}{D_n} \sum_{\zeta \in \mathcal{D}_n} l ( \mathbf{w}, \zeta).
	\end{align}
	
	Ultimately, FL aims to derive the optimal global model that minimizes the global loss function
	\begin{align}
		\mathbf{w}^* = \argmin_{\mathbf{w} \in \mathbb{R}^m} f(\mathbf{w} ).
	\end{align}
	
	In general, FL iteratively follows the procedures of local training, local update transmission, global model update, and global model distribution.
	First, in the local training, IoT devices perform stochastic gradient descent based on its local dataset. 
	We denote a set of IoT devices performing local training in training round $k$ as $\mathcal{C}_k$. 
	Based on the most recently received global model, IoT device $n\in\mathcal{C}_k$ computes gradient using a mini-batch $\mathcal{B}_{n,k}$ of data samples randomly selected from its local dataset. The mini-batch size is assumed to be identical to all IoT device and fixed across all training rounds, $|\mathcal{B}_{n,k}| = B$.
	Suppose each IoT device performs $H$ stochastic gradient step.
	Given that time is slotted and IoT devices are capable of processing $q$ data samples per time slot, the time duration for local training becomes
	\begin{align}
		\taucomp = \left\lceil \frac{H B}{q} \right\rceil. \label{eq:t_comp}
	\end{align}
	
	When IoT devices complete local training, they are ready to transmit the local updates. 
	The set of IoT devices ready for transmission in training round $k$ is denoted as $\mathcal{N}^{\text{avail}}_k$. 
	
	TDMA is considered for transmitting local updates in bandwidth-limited scenarios. In each time slot, an IoT device from $\mathcal{N}^{\text{avail}}_k$ will transmits its local updates. If all IoT devices are not ready for transmission, the server waits until any device completes its training. On the other hand, when multiple IoT devices are ready for transmission, the device which finishes local computation earliest transmits, breaking ties based on their indexes.
	
	In each training round, the server receives local updates from $S$ IoT devices via TDMA to update the global model.
	The set of IoT devices which send local updates in training round $k$ is denoted as $\mathcal{S}_k\in\mathcal{N}^{\text{avail}}_k$.
	
	After receiving the local updates, the server updates the global model using the aggregated local update.
	Let $\bar{\mathbf{g}}_k$ denote the aggregated local update in training round $k$. 
	Then, the global model is updated as
	\begin{align}
		\mathbf{w}_{k+1} = \mathbf{w}_k - \eta \bar{\mathbf{g}}_k, \label{eq:sgd}
	\end{align} 
	where $\eta$ is the step size.	Then, the updated global model is sent back to IoT devices which transmitted their local update for the next training round.Due to the limited capabilities of IoT devices, It is infeasible for IoT devices to compute and communicate simultaneously.
	
	Suppose $r$ time slots are required for each IoT device to transmit its local update and for the server to send the updated global model. 	Since $S$ IoT devices and the server transmit in every training round, the number of time slots for communication in each training round is calculated as
	\begin{align}
		\taucomm = r(S + 1). \label{eq:t_comm}
	\end{align}
	
	By repeating the procedures from local training to the distribution of the global model, the global model is trained by leveraging distributed local datasets. 
	However, in practical scenarios,  we have a deadline at which we need to finish training. Thus, we focus on practical FL that aims to derive the global model achieving the lowest loss within a given time window. 
	Let $T$ be the number of time slots assigned for training.	
	
	\section{Asynchronous Federated Learning Over TDMA Channel}\label{sec:asynch}
	
	In a vanilla  FL setup, the local training of devices begins from the same global model  which the server delivers at the end of the previous training round. If the global model in training round $k$ is represented as $\mathbf{w}_k$, the aggregated local update in training round $k$ is represented as 
	\begin{align}
		\bar{\mathbf{g}}_k= \frac{1}{S} \sum_{n\in\mathcal{S}_k} \nabla f_n\left( \mathbf{w}_k ; \mathcal{B}_{n,k} \right), \label{eq:syn_AggUpdate}
	\end{align}
	where $\mathcal{B}_{n,k}$ denotes the mini-batch of device $n$ in training round $k$.   
	Due to synchronicity, computations and communications can be strictly divided. Hence, the number of time slots for each training round in synchronous FL becomes
	\begin{align}
		\tau^\text{syn} = \tau^\text{comp} + \tau^\text{comm}.
	\end{align}
	Since all devices perform local training simultaneously, no transmissions occur during these time slots, leading to a waste of communication resources.	Furthermore, this inefficiency becomes more severe in scenarios where local training times are prolonged due to the low computational power of IoT devices.
	
	To mitigate this inefficiency in practical systems, we propose asynchronous FL, which fully utilizes wireless channels through the introduction of asynchronous local updates.
	
	Specifically, while IoT device $n\in\mathcal{S}_{k}$ communicates with the server, the other devices can compute their local updates. Consequently, in asynchronous FL, computations and communications can be carried out within the same time slot by different groups of devices. Moreover, time duration for a training round in asynchronous FL can be different depending on amount of overlap between computations and communications. Let $G=\left\lceil\frac{N}{S}\right\rceil$ denote the number of TDMA groups. Then, if $\taucomp$ is greater than or equal to communication time for $G-1$ TDMA groups, the time duration of $G$ training rounds consists of the communication time for all $G$ TDMA groups, $r(s+1)G$, plus the additional computation time not overlapped with communication, $\taucomp - r(S+1)(G-1)$. Consequently, the time slots for $G$ training round becomes 
		\begin{align}
			\tau^{G \text{ rounds}}	&=\taucomp + r(S+1).
	\end{align}
	On the other hand, if $\taucomp$ is less than communication time for $G-1$ TDMA groups, communication time is enough to hide computation time. Thus, time duration of $G$ training round becomes communication time of $G$ TDMA groups.  Consequently, the average number of time slots for a training round in asynchronous FL over TDMA channel is given as follows.
	\begin{align}\label{eq:tau_asyn}
		\tau^\text{asyn}=\begin{cases}
			\frac{\taucomp + r (S+1)}{G} & \text{ if } \taucomp \geq r \left(G - 1 \right) (S + 1) \\
			\tau^{\text{comm}} & \text{ otherwise }
		\end{cases} .
	\end{align}
	
	However, in asynchronous FL, global model used for computing local updates can be different. In training round $k$, the aggregated local update in asynchronous FL is represented as
	\begin{align}
		\bar{\mathbf{g}}_k = \frac{1}{S} \sum_{n \in \mathcal{S}_k} \nabla f_n(\mathbf{w}_{\hat{k}_n}; \mathcal{B}_{n, 
			\hat{k}_n}),
	\end{align} where $\hat{k}_n$ denotes the training round when IoT device $n$ starts local training.
	Specifically, we can express $\hat{k}_n$ as 
	\begin{align}
		\hat{k}_n = \begin{cases}
			0 &  \left\lbrace  k' < k| n \in \mathcal{S}_{k'} \right\rbrace = \emptysetAlt \\ 
			\max \left\lbrace  k' < k| n \in \mathcal{S}_{k'} \right\rbrace + 1 & \text{ otherwise}
		\end{cases}, \label{eq:def_k_tilde}
	\end{align} where $\emptysetAlt$ is an empty set. Note that the condition $\left\lbrace  k' < k| n \in \mathcal{S}_{k'} \right\rbrace = \emptyset$ is equivalent that device $n$ has not transmitted to the server before training round $k$.
	
	Using $\bar{\mathbf{g}}_k$, the server updates the global model and the updated model $\mathbf{w}_{k+1}$ is distributed to all IoT devices in $\mathcal{S}_k$.	
	
	\begin{figure*}[!htb]
		\centering
		\includegraphics[scale=0.5]{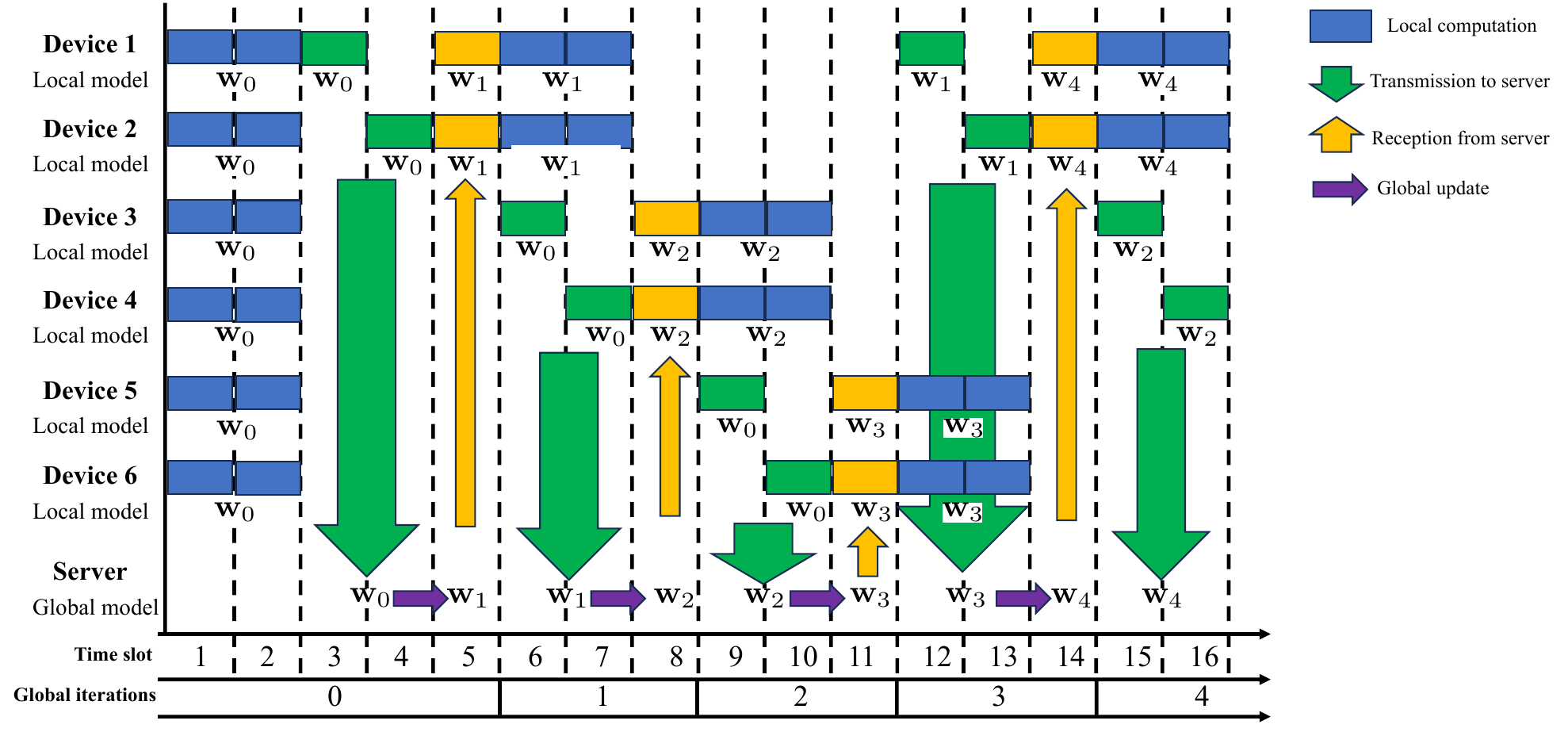}
		\caption{Example timeline of asynchronous FL over TDMA channel for $N=6, S = 2, \taucomp = 1$, and $r=1$ }\label{fig:example}
	\end{figure*}
	
	For example, consider a scenario where $N=6$ and $S = 2$, $\taucomp = 2$ and $r=1$ as illustrated in Fig. \ref{fig:example}.
	The asynchronous FL begins with the local training of all IoT devices using the initial global model $\mathbf{w}_0$. 
	After $\taucomp=2$ time slots, all devices complete the local training and are ready to transmit the local updates to the server.
	Hence, $\mathcal{N}^{\text{avail}}_0 = \mathcal{N}$. Moreover, $\hat{k}_n = 0$ for $n \in  \mathcal{N}^{\text{avail}}_0$.
	Among the devices ready to transmit, $S=2$ devices actually send their local updates using TDMA scheme in training round $k=0$. It is supposed for IoT devices which complete local computation earlier to send their local updates. However, for training round $0$, as all IoT devices has finished local updates simultaneously, ties are broken by selecting lower index. Thus, the IoT devices in $\mathcal{S}_0=\{1, 2\}$ send their local updates $\nabla f_1(\mathbf{w}_0, \mathcal{B}_{1,0})$ and $\nabla f_2(\mathbf{w}_0; \mathcal{B}_{2,0}))$ to the server at the time slots 2 and 3, respectively.
	Based on the aggregated local updates from devices in $\mathcal{S}_0$, the global model is updated as  
	\begin{align}
		\mathbf{w}_1 = \mathbf{w}_0 -  \frac{\eta}{2}  \left( \nabla f_1 (\mathbf{w}_0 ;\mathcal{B}_{1,0}) + \nabla f_2 (\mathbf{w}_0; \mathcal{B}_{2,0}) 
		\right).
	\end{align}
	The server then sends $\mathbf{w}_1$ to the devices in $\mathcal{S}_0$ at time slot 4, marking the end of training round $k=0$. 
	
	From time slot $5$, the devices in $\mathcal{S}_0$ start to compute local updates again based on $\mathbf{w}_1$. 
	Concurrently, IoT devices in $\mathcal{S}_1={3, 4}$ transmit their local updates, $\nabla f_3(\mathbf{w}_0, \mathcal{B}_{3,0})$ and $\nabla f_4(\mathbf{w}_0, \mathcal{B}_{4,0})$, during time slots 6 and 7, respectively.
	Based on the received local updates from devices in $\mathcal{S}_1$, $\mathbf{w}_1$ is updated as
	\begin{align}
		\mathbf{w}_2 = \mathbf{w}_1 - \frac{\eta}{2}  \left( \nabla f_3 (\mathbf{w}_0 ;\mathcal{B}_{3,0}) + \nabla f_4 (\mathbf{w}_0; \mathcal{B}_{4,0}) 
		\right ).
	\end{align}
	Then, the server sends $\mathbf{w}_2$ back to devices in $\mathcal{S}_1$ during time slot 7.   
	
	Note that the local updates used to update the global model in training round 1 are generated based on outdated global model $\mathbf{w}_0$, not the up-to-date model $\mathbf{w}_1$.   
	Similarly, in training round 2, $\mathbf{w}_3$ is generated using outdated local updates from IoT devices in $\mathcal{S}_2=\{5, 6\}$.
	However, the outdatedness of the update increases to 2 rounds in this training round.
	In other words, except for the first training round, local training is conducted based on the outdated global model when $S<N$.
	
	\begin{algorithm}[t!]
		\caption{Asynchronous Federated Learning over TDMA Channel}\label{alg:async}
		\begin{algorithmic}[1]
			\renewcommand{\algorithmicrequire}{\textbf{Input:}}
			\newcommand{\algorithmicforp}{\algorithmicfor \textbf{ do in parallel}}
			\State Input: $\mathcal{N}$, $\mathcal{D}_n$, $\eta$, $\mathbf{w}_0$, $\mathcal{S}_{- 1} = \{ 1, 2, \dots, N \}$ 
			%\ENSURE  out
			\For {$k \in \{ 0,1,2,\dots\ \lfloor T/\tau^\text{asyn}\rfloor -1 \}$}
			\For {$n \in \mathcal{S}_{k - 1}$} in parallel
			\State Update $\hat{k}_n \leftarrow k$
			\State Start computing local update $\nabla f_n ( \mathbf{w}_k ; \mathcal{B}_{n,k})$			
			\EndFor
			\State Update $\mathcal{N}^{\text{avail}}_{k}$
			\State $\mathcal{S}_{k}\leftarrow \underset{\mathcal{N}' \subseteq \mathcal{N}^{\text{avail}}_{k} , |\mathcal{N}'| = S }{\argmin}  
			\sum_{n 
				\in 
				\mathcal{N}'} \hat{k}_n$
			\For {$n \in \mathcal{S}_k$} 
			\State Transmit $\nabla f_n(\mathbf{w}_{\hat{k}_n}; \mathcal{B}_{n, \hat{k}_n})$ using TDMA
			\EndFor
			\State $\mathbf{w}_{k+1} = \mathbf{w}_k -\frac{ \eta }{S} \sum_{n \in \mathcal{S}_k} \nabla  f_n(\mathbf{w}_{\hat{k}_n} ; 
			\mathcal{B}_{n , \hat{k}_n})$
			\State Server sends $\mathbf{w}_{k+1}$ to device $n \in \mathcal{S}_{k}$             
			\EndFor
		\end{algorithmic}
	\end{algorithm}
	
	The overall procedure of the asynchronous FL over TDMA channel is summarized in Algorithm \ref{alg:async} for general scenario.
	It is worth noting that procedures from lines 3 to 6 and 7 to 13 in Algorithm \ref{alg:async} are conducted concurrently. 
	While some IoT devices compute local updates, others can communicate with the server. 
	
	Let us define the staleness of local update that IoT device $n\in \mathcal{S}_k$ transmitted to the server as
	\begin{align}
		d_{n,k} \delequal k - \hat{k}_n. \label{eq:delay_0}
	\end{align}
	
	Suppose $S = N$. 
	In this case, all IoT devices transmit in every training round, leading to $\hat{k}_n = k$.
	As a result, $d_{n,k} = 0$ for any $n$ and $k$, which implies that FL becomes synchronous.
	On the other hand, if $S = 1$, only one of IoT devices can transmit its local update in each training round.
	Hence, the staleness of local update is maximized since each IoT device can only transmit its local update every $N$ rounds.
	
	In fact,  $d_{n,k}$ depends on $S$. 
	This dependency can be represented as 
	\begin{align}\label{eq:delay}
		d_{n,k} = \begin{cases}
			k & \text{ for }  k < G  \\
			G - 1,  & \text{ for } k \geq G
		\end{cases}
	\end{align} 
	In \eqref{eq:delay}, $d_{n,k}$ for $k < G$ arises from the fact that when more than $S$ devices have the same $\hat{k}_n$, those with lower indices are selected for transmission. 
	On the other hand, $d_{n,k}$ becomes identical as $G-1$ for training round $k\le G$.
	
	Based on \eqref{eq:delay}, the update of the global model at the server is expressed as 
	\begin{align}		
		\mathbf{w}_{k+1} =\begin{cases}
			\mathbf{w}_k - \frac{\eta}{S} \sum\limits_{n \in \mathcal{S}_k} \nabla 
			f_n(\mathbf{w}_0; 
			\mathcal{B}_{n, k}), & \text{for } k < G
			\\
			\mathbf{w}_k - \frac{\eta}{S} \sum\limits_{n \in \mathcal{S}_k} \nabla 
			f_n(\mathbf{w}_{k-G+1}; 
			\mathcal{B}_{n, k}), & \text{for } k\ge G
		\end{cases} .
	\end{align}

	\section{Convergence Analysis}\label{sec:convergence}
	In this section, we present the convergence analysis of asynchronous FL over TDMA channel.	
	Before proving the convergence, we first state the assumptions that widely used for convergence analysis of FL.
	\begin{assumption} \label{assump_smooth}
		The global loss function $f(\cdot)$ is $L$-smooth function if it satisfies the following inequality for any $\mathbf{w}$ and $\mathbf{w}'$: 
		\begin{align}
			\Vert \nabla f( \mathbf{w}) - \nabla f( \mathbf{w}')\Vert \leq L \Vert \mathbf{w} - \mathbf{w}'\Vert. \label{ineq:smooth}
		\end{align}
		Moreover, \eqref{ineq:smooth} implies that
		\begin{align}
			f( \mathbf{w}' ) \leq f(\mathbf{w}) + \nabla( f(\mathbf{w}))^\intercal ( \mathbf{w}'- \mathbf{w}) + \frac{L}{2} \Vert \mathbf{w}'- \mathbf{w}\Vert^2. 
		\end{align}
	\end{assumption}
	\begin{assumption}\label{assump_bound}
		The second moment of local loss function for any $n$ and $\mathbf{w}$ is bounded above by
		\begin{align}
			\mathbb{E} \left[ \Vert \nabla f_n ( \mathbf{w} ; \zeta ) \Vert^2 \right] \leq \sigma^2 + M \Vert \nabla f_n(\mathbf{w}) \Vert^2,
		\end{align} where $\zeta$ is a data sample.
	\end{assumption}
	Under the assumption \ref{assump_bound}, we can obtain similar bound for mini-batch gradient.
	\begin{align}
		\mathbb{E} \left[ \| \nabla f_n (\mathbf{w} ; \mathcal{B}_{n}) \|^2 \right] \leq \frac{\sigma^2}{B} + \frac{M}{B} \| \nabla f_n (\mathbf{w} ) \|^2.
	\end{align}
	\begin{assumption}\label{assump_hetero}
		We assume that the heterogeneity of data distribution is bounded. For any $n$ and $\mathbf{w}$, 
		\begin{align}
			\Vert \nabla f(\mathbf{w}) - \nabla f_n(\mathbf{w}) \Vert^2 \leq \Gamma^2. \label{ineq:assump_hetero}
		\end{align}
	\end{assumption}
	
	Based on the above assumptions, we first prove the following lemma which states the upper 
	bound of the improvement of loss of global model.
	
	\begin{lemma}\label{lem:descent_1}
		Under Assumptions \ref{assump_smooth}, \ref{assump_bound}, and \ref{assump_hetero}, we can bound $\mathbb{E} \left[ f(\mathbf{w}_{k+1} ) \right] $ as follows
		\begin{align}
			\nonumber &\mathbb{E} \left[ f(\mathbf{w}_{k+1} ) \right] \leq  \mathbb{E}[ f(\mathbf{w}_k)]  - \frac{\eta}{2} 
			\mathbb{E} \left[ \Vert \nabla 
			f(\mathbf{w}_k) \Vert^2 \right] \\
			\nonumber & + \left( \frac{\eta^2 M L}{2 S^2 B}-\frac{\eta}{2S} \right) \sum_{n \in 
				\mathcal{S}_k} \mathbb{E} \left[ 
			\left\| \nabla f_n (\mathbf{w}_{\hat{k}_n} ) \right \|^2 \right] + \frac{\eta \Gamma^2}{2} \\
			& + \frac{\eta L^2}{2 S} \sum_{n \in \mathcal{S}_k} \mathbb{E} \left[ \left \| \mathbf{w}_k - 
			\mathbf{w}_{\hat{k}_n} \right \|^2 \right]  + \frac{\eta^2 	\sigma^2  L }{2 S B}. \label{ineq:descent}
		\end{align}	
		\begin{IEEEproof}
			Refer to Appendix \ref{App:lemma1}.
		\end{IEEEproof}
	\end{lemma}
	
	Furthermore, In the following lemmas, we obtain bounds for terms in \eqref{ineq:descent}. 
	\begin{lemma}\label{lem:OutdatedW}
		Under Assumption \ref{assump_hetero}, we have
		\begin{align}
			\sum_{n\in\mathcal{S}_k} \mathbb{E}\!\left[ \Vert f(\mathbf{w}_{\hat{k}_n}) \Vert^2 \right] \le  \Gamma^2S + \sum_{n\in\mathcal{S}_k}\mathbb{E}\!\left[ \Vert \nabla f(\mathbf{w}_{\hat{k}_n}) \Vert^2 \right]   \label{ineq:conv6}
		\end{align}
		\begin{IEEEproof}
			Based on triangular inequality and Assumption \ref{assump_hetero}, 
			\begin{align}
				&\sum_{n \in \mathcal{S}_k} \mathbb{E} \left[ \Vert \nabla f_n ( \mathbf{w}_{\hat{k}_n}) \Vert ^2 \right] \nonumber\\
				& = \sum_{n \in \mathcal{S}_k} \mathbb{E} \left[ \Vert \nabla f_n(\mathbf{w}_{\hat{k}_n}) - \nabla f(\mathbf{w}_{\hat{k}_n}) + 
				\nabla f (\mathbf{w}_{\hat{k}_n}) \Vert^2 \right] , \nonumber\\
				& \leq \sum_{n \in \mathcal{S}_k} \mathbb{E} \left[ \Vert \nabla f_n(\mathbf{w}_{\hat{k}_n}) - \nabla f(\mathbf{w}_{\hat{k}_n}) 
				\Vert^2 + \Vert \nabla f (\mathbf{w}_{\hat{k}_n}) \Vert^2 \right] , \nonumber\\
				& \leq  \Gamma^2 S + \sum_{n \in \mathcal{S}_k} \mathbb{E} \left[  \Vert \nabla f (\mathbf{w}_{\hat{k}_n}) 
				\Vert^2 \right] . \nonumber
			\end{align}            
		\end{IEEEproof}
	\end{lemma}
	
	\begin{lemma}\label{lem:lemma3}
		Under Assumption \ref{assump_hetero}, $\sum_{n\in\mathcal{S}_k} \mathbb{E}\!\left[ \Vert \mathbf{w}_k-\mathbf{w}_{\hat{k}_n} \Vert^2 \right] $ can be bounded as follows.
		\begin{align}
			\sum_{n\in\mathcal{S}_k} \mathbb{E}\!\left[ \Vert \mathbf{w}_k-\mathbf{w}_{\hat{k}_n} \Vert^2 \right] &\le \frac{\eta^2(\sigma^2+M\Gamma^2)}{SB} \sum_{n\in\mathcal{S}_k} d_{n,k} \nonumber\\
			&~~~+\frac{\eta^2M}{S^2B} \sum_{n\in\mathcal{S}_k}\sum_{j=\hat{k}_n}^{k-1}\sum_{n'\in\mathcal{S}_j} \mathbb{E}\!\left[ \Vert \nabla f(\mathbf{w}_{\hat{j}_{n'}}) \Vert^2 \right]
		\end{align}
		\begin{IEEEproof}
			Refer to Appendix \ref{App:lemma3}
		\end{IEEEproof}
	\end{lemma}
	
	Using the result of Lemma \ref{lem:OutdatedW} and Lemma \ref{lem:lemma3}, the inequality \eqref{ineq:descent} can be further bounded above as
	\begin{align}
		&\mathbb{E} \left[ f(\mathbf{w}_{k+1} ) \right] \nonumber\\
		&\leq  \mathbb{E}[ f(\mathbf{w}_k)]  - \frac{\eta}{2} \mathbb{E} \left[ \Vert \nabla f(\mathbf{w}_k) \Vert^2 \right] \nonumber \\
		&~~~+ \frac{\eta^2 M L - \eta SB}{2 S^2 B}  \left( \Gamma^2 S + \sum_{n \in 
			\mathcal{S}_k} \mathbb{E} \left[ \left\| \nabla f(\mathbf{w}_{\hat{k}_n} )\right \|^2 \right] \right) + \frac{\eta \Gamma^2}{2} \nonumber \\
		&~~~ + \frac{\eta L^2}{2 S} \left( \frac{\eta^2 \left(\sigma^2 + M \Gamma^2 \right) }{S B } \sum_{n \in \mathcal{S}_k} d_{n,k} \right. \nonumber \\
		&\left.~~~ + \frac{\eta^2 M}{S^2 B } \sum_{n \in \mathcal{S}_k} \sum_{j= \hat{k}_n}^{k-1}  \sum_{n' \in \mathcal{S}_j} \mathbb{E} \left[  \Vert \nabla f (\mathbf{w}_{\hat{j}_{n'}} ) 
		\Vert^2 \right]  \right)  + \frac{\eta^2 	\sigma^2  L }{2 S B}, \nonumber\\
		& = \mathbb{E}[ f(\mathbf{w}_k)]  - \frac{\eta}{2} \mathbb{E} \left[ \Vert \nabla 
		f(\mathbf{w}_k) \Vert^2 \right] +  \frac{\eta^2 L \left( \Gamma^2 M +  \sigma^2 \right) }{2 S B}\nonumber   \\
		&~~~+  \frac{\eta^2 M L - \eta SB }{2 S^2 B}   \sum_{n \in 
			\mathcal{S}_k} \mathbb{E} \left[ \left\| \nabla f(\mathbf{w}_{\hat{k}_n} )\right \|^2 \right] \nonumber \\				
		&~~~ + \frac{\eta^3 L^2 M}{2S^3 B } \sum_{n \in \mathcal{S}_k} \sum_{j= \hat{k}_n}^{k-1}  
		\sum_{n' \in 
			\mathcal{S}_j} \mathbb{E} \left[  \Vert \nabla f (\mathbf{w}_{\hat{j}_{n'}} ) 
		\Vert^2 \right] \nonumber \\
		&~~~ + \frac{\eta^3 L^2 \left(\sigma^2 + M \Gamma^2 \right) }{2 S^2 B } \sum_{n \in \mathcal{S}_k
		} d_{n,k}. \label{ineq:descent_2}
	\end{align}
	
	By summing \eqref{ineq:descent_2} from $k=0$ to $K$, and dividing by $K+1$, we obtain
	\begin{align}
		\nonumber & \frac{\eta}{K+1} \sum_{k=0}^K \mathbb{E} \left[ \left \| \nabla f(\mathbf{w}_k) \right\|^2 \right] \\
		&- \frac{\eta^2 M L - \eta S B}{ 
			S^2 B (K+1)}\sum_{k=0}^K \sum_{n \in \mathcal{S}_k} \mathbb{E} \left[ \left\| \nabla f(\mathbf{w}_{\hat{k}_n} )\right 
		\|^2 \right] \nonumber  \\
		\nonumber & -  \frac{\eta^3 L^2 M}{S^3 B (K+1)}\sum_{k=0}^K 
		\sum_{n \in \mathcal{S}_k} \sum_{j= \hat{k}_n}^{k-1}  \sum_{n' \in \mathcal{S}_j} \mathbb{E} \left[  \Vert 
		\nabla f (\mathbf{w}_{\hat{j}_{n'}} ) \Vert^2 \right] \\
		& \leq  \frac{2}{K+1}  \mathbb{E}\left[ f(\mathbf{w}_0)  -  f(\mathbf{w}_{k+1} )\right]  + \frac{\eta^2 L \left( \Gamma^2 M +  
			\sigma^2 \right) }{ S B}  \nonumber \\
		& + \frac{\eta^3 L^2 \left(\sigma^2 + M \Gamma^2 \right) }{S^2 B (K+1) }\sum_{k=0}^K \sum_{n \in 
		} d_{n,k}. \label{ineq:descent_3}
	\end{align}
	
	Assuming that the training round is greater than twice of the number of TDMA groups, $K\ge 2 G$, the terms associated with the outdated local update in \eqref{ineq:descent_3} can be simplified as described in the following lemma.
	\begin{lemma}\label{lem:sq_norm_dgrad}
		For $K\ge 2 G$, the sums of the squared norms of the outdated local updates are bounded above as
		\begin{align}
			\nonumber \sum_{k=0}^{K} \sum_{n \in \mathcal{S}_k} &\mathbb{E} \left[ \Vert \nabla f(\mathbf{w}_{\hat{k}_n}) \Vert^2 \right] \leq (N - 
			S)  \Vert \nabla f(\mathbf{w}_0) \Vert^2  \\
			& + S  \sum_{k=0}^K \mathbb{E} \left[ \Vert \nabla f(\mathbf{w}_k) \Vert^2 \right] \label{eq:conv1_10} , 
		\end{align} and 
		\begin{align}
			&\sum_{k=0}^K \sum_{n \in\mathcal{S}_k} \sum_{j = \hat{k}_n}^{k-1} \sum\limits_{n' \in \mathcal{S}_j} \mathbb{E} 
			\left[ \Vert \nabla f (\mathbf{w}_{\hat{j}_{n'}} ) \Vert ^2 \right] \nonumber \\
			& \leq 2 G S^2 \left( G \left[ \| \nabla f(\mathbf{w}_0)\|^2 \right] + \sum_{k=0}^{K} \mathbb{E} \left[ \| \nabla 
			f(\mathbf{w}_k)\|^2\right] \right) .
			\label{ineq:conv1_9}
		\end{align}	
		\begin{IEEEproof}
			Refer to Appendix \ref{App:lemma4}.
		\end{IEEEproof}
	\end{lemma}
	
	By incorporating the results of Lemma \ref{lem:sq_norm_dgrad} into inequality \eqref{ineq:descent_3}, we prove the convergence of asynchronous FL over a TDMA channel, as stated in Theorem \ref{thm:convergence}.
	
	\begin{theorem}\label{thm:convergence}
		Suppose that $N$ is divisible by $S$ and the stepsize $\eta$ satisfies  that
		\begin{align}
			\eta \leq  \frac{\beta}{\sqrt{K+1}}, 
		\end{align} where $\beta$ is defined as $\beta = \frac{S B}{2LN} \left( \sqrt{ 1 + \frac{8N}{BM}} - 1 
		\right)$. Then, under Assumptions \ref{assump_smooth}, 
		\ref{assump_bound}, and \ref{assump_hetero},  we have
		\begin{align}
			\frac{1}{K+1} \sum_{k=0}^K \mathbb{E} \left[ \Vert \nabla f (\mathbf{w}_k) \Vert^2 \right] =\mathcal{O} \left(   
			\frac{G^2}{\sqrt{K}}   \right).  
		\end{align}		
		\begin{IEEEproof}
			Refer to Appendix \ref{App:theorem1}.
		\end{IEEEproof}
	\end{theorem}

	\section{Intentional Delay Federated Learning}\label{sec:int_delay}	
	
	From the result of Theorem \ref{thm:convergence}, the convergence of asynchronous FL 
	over TDMA channel depends on the number of communication groups. Since the staleness of local updates is given by $G-1$ for training rounds $k\ge G$, a large number of groups implies that local updates are significantly outdated.
	Intuitively, a larger delay suggests that the model used to compute the stochastic gradient is likely to differ more from the model to which the gradient is applied. As a result, the stochastic gradient computed from a past global model may deviate significantly from the true gradient of the current global model, with the extent of this deviation increasing as $G$ becomes larger.
	
	Since $G$ is equal to $\frac{N}{S}$, increasing the TDMA group size reduces the outdatedness.     However, since a training round ends only after $S$ IoT devices have sequentially transmitted their local updates, the number of time slots required for a singe update of global model can become excessive when $S$ is large.
	Therefore, in practical scenarios with limited training time, it is important to balance both reducing the outdatedness of local updates and increasing the frequency of global model updates.
	
	\begin{algorithm}[t!]
		\caption{Intentional Delay Federated Learning}\label{alg:IDFL}
		\begin{algorithmic}[1]
			\renewcommand{\algorithmicrequire}{\textbf{Input:}}
			\newcommand{\algorithmicforp}{\algorithmicfor \textbf{ do in parallel}}
			\State Input: $\mathcal{N}$, $\mathcal{D}_n$, $\eta$, $\mathbf{w}_0$, $ \mathcal{S}_{-\alpha-1} = \{  1, \dots, ( G-\alpha  )S \} $, $ \mathcal{S}_{k-\alpha-1} = \{ ( k + G - \alpha - 1)S + 1, \dots, (k + G-\alpha )S \} $ for $1 \leq k \leq \alpha $.
			%\ENSURE  out
			\For {$k \in \{ 0,1,2,\dots\ \lfloor T/\tau^\text{asyn}\rfloor -1 \}$}
			\For {$n \in \mathcal{S}_{k - \alpha - 1}$} in parallel
			\State Update $\hat{k}_n \leftarrow k$
			\State Start computing local update $\nabla f_n ( \mathbf{w}_k ; \mathcal{B}_{n,k})$			
			\EndFor
			\State Update $\mathcal{N}^{\text{avail}}_{k}$
			\State $\mathcal{S}_{k}\leftarrow \underset{\mathcal{N}' \subseteq \mathcal{N}^{\text{avail}}_{k} , |\mathcal{N}'| = S }{\argmin}  
			\sum_{n \in \mathcal{N}'} \hat{k}_n$
			\For {$n \in \mathcal{S}_k$} 
			\State Transmit $\nabla f_n(\mathbf{w}_{\hat{k}_n}; \mathcal{B}_{n, \hat{k}_n})$ using TDMA
			\EndFor
			\State $\mathbf{w}_{k+1} = \mathbf{w}_k -\frac{ \eta }{S} \sum_{n \in \mathcal{S}_k} \nabla  f_n(\mathbf{w}_{\hat{k}_n} ; 
			\mathcal{B}_{n , \hat{k}_n})$
			\State Server sends $\mathbf{w}_{k+1}$ to device $n \in \mathcal{S}_{k-\alpha}$             
			\EndFor
		\end{algorithmic}
	\end{algorithm}
	
	To achieve faster and more stable learning, we propose intentional delay FL (IDFL) which introduces an additional delay to receive a more recent global model before computing the gradient, as described in Algorithm \ref{alg:IDFL}.
	In conventional FL, each device receives the updated global model at the end of the same training round in which the device transmitted local update.
	On the other hand, in IDFL, each device waits for a predefined number of training rounds, referred to as an intentional delay to reduce the staleness of local update by receiving a more recent global model. Specifically, the IoT devices that transmitted their local updates in training round $k$ receive the updated global model in training round $k+\alpha$.
    Then, \eqref{eq:def_k_tilde} can be rewritten as 
    \begin{align}
		\hat{k}_n = \begin{cases}
			0 &  k\le G-\alpha \\ 
                k-G+\alpha & G-\alpha<k\le G  \\
			\max \left\lbrace  k' < k| n \in \mathcal{S}_{k'} \right\rbrace +\alpha+ 1 & \text{ otherwise}
		\end{cases}.
	\end{align}
	
	To minimize the staleness of local updates while preserving the parallel execution of computation and communication in asynchronous FL, the intentional delay needs to be carefully chosen. The following proposition provides guidance on selecting the intentional delay that minimizes the staleness of local updates without compromising the benefits of asynchronous FL.
	\begin{proposition}\label{prop:max_idelay}
		Assuming $N$ is divisible by $S$, the intentional delay $\alpha$ which minimizes the staleness of local updates without increasing the number of time slots for each training round is given by
		\begin{align}
			\alpha &= \begin{cases}
				0 & \text{ if } \frac{\taucomp}{r} \geq (G-1)(S+1) \\
				G - d^*  - 1& \text{ if } \frac{\taucomp}{r}  < (G-1)(S+1)
			\end{cases} , \label{eq:max_delay}
		\end{align} where $d^*$ is an integer which satisfies
		\begin{align}
			(d^* -1)(S+1) < \frac{\taucomp}{r} \leq d^* (S+1). \label{ineq:ratio}
		\end{align}
		
		\begin{IEEEproof}
			As $N$ is divisible by $S$, we have $N=GS$.
			Suppose IoT devices in group $g$ transmit in training round $k$. 
			Given the length of training round \eqref{eq:tau_asyn}, the devices in group $g$ need to wait at least $r(G-1)(S+1)$ time slots before their next local update transmission.
			
			In particular, for $\taucomp \geq r(G-1)(S+1)$, the local update of IoT devices in group $g$ cannot be 
			completed even after all the other groups have completed their local update transmissions.
			Since a non-zero intentional delay leads to  additional waiting time for the completion of local training, it is desirable to set it to zero.
			
			On the other hand, if $\taucomp < r(G-1)(S+1)$, it is feasible for IoT devices to avoid the extra time slots for the completion of local training when intentional delay is applied.	Suppose IoT devices in group $g$ send local updates in training round $k$. 	Then, the next training round that group $g$ communicates is training round $k + G - 1$. 
			To avoid waste of time slot, time interval between training round $k + \alpha$ and $k + G - 1$  
			should be greater than or equal to the training time of IoT devices as follows
			\begin{align}
				r( G - \alpha -1 ) (S + 1)\ge \taucomp. \label{ineq:max_d_1}
			\end{align}
			
			On the other hand, if local update is completed before the beginning of training round $k + G -1$ , it is allowed to wait more time slots for devices without extra time slots for local training completion.
			Consequently, when the maximum intentional delay is applied, IoT devices need to finish computation right before they transmit. Hence, the time slots between training round $k +G-2$ and $k + \alpha$  should be less than the computing time.
			\begin{align}
				\taucomp > r ( G - \alpha -2) (S+1). \label{ineq:max_d_2}
			\end{align} 
			
			Consequently, based on  \eqref{ineq:max_d_1} and \eqref{ineq:max_d_2}, we have
			\begin{align}\label{eq:taucomp_1}
				r ( G - \alpha -2) (S+1)  < \taucomp \leq r( G - \alpha -1 ) (S + 1) .
			\end{align}
			We define an effective delay in IDFL as $d^*=G-\alpha-1$ as $d^*$.
			The inequality \eqref{eq:taucomp_1} can be rewritten as \eqref{ineq:ratio}. 
			Combining the case when $\taucomp \geq r (G-1) (S+1)$, we have \eqref{eq:max_delay}.
		\end{IEEEproof}
	\end{proposition}

	When intentional delay is applied, the staleness of local updates are reduced to $d^*=G - 1 - \alpha$.	
	As $\taucomp$ and $r$ represent the time slots spent on local training and model delivery, respectively, $\frac{\taucomp}{r}$ can be used to parameterize the trade-off between computation and communication. 
	If $\frac{\taucomp}{r}$ is large, Proposition \ref{prop:max_idelay} shows that intentional delay cannot be exploited. 
	Intuitively, when each device have low computing capability, it is required to start early to be ready for transmission in right 
	time. 
	On the other hand, if communication burden is large, appropriate waiting is allowed since each device has enough time until to be selected for 
	next transmission.
	
	\section{Experiments}\label{sec:exp}
	
	\begin{figure*}[t!]
		\centering
		\subfloat[Global loss for MNIST dataset]{\label{fig:config_1_0_loss}
			\includegraphics[scale=0.35]{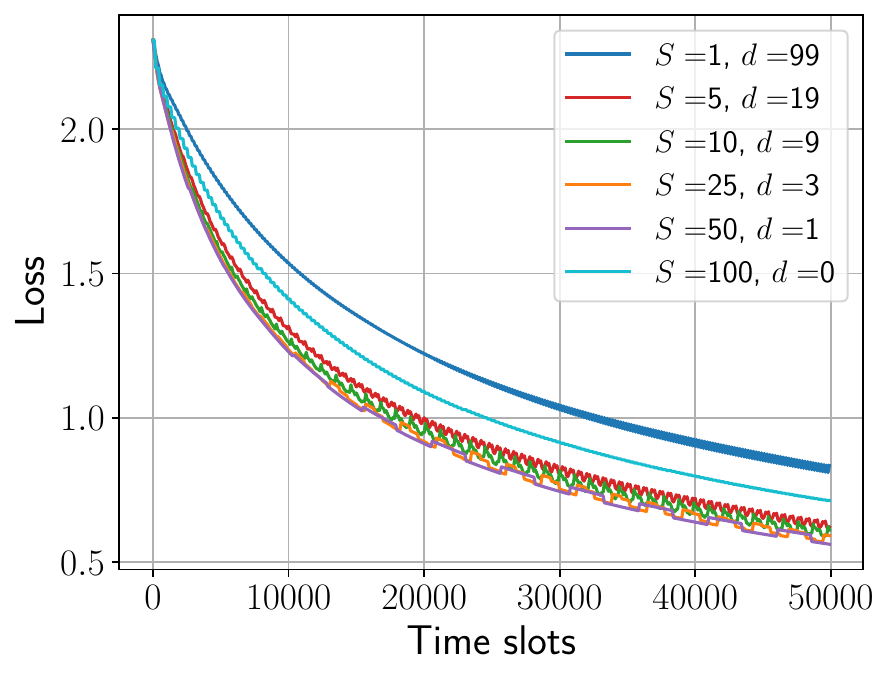}			
		}
		\subfloat[Test accuracy for MNIST dataset]{\label{fig:config_1_0_acc}
			\includegraphics[scale=0.35]{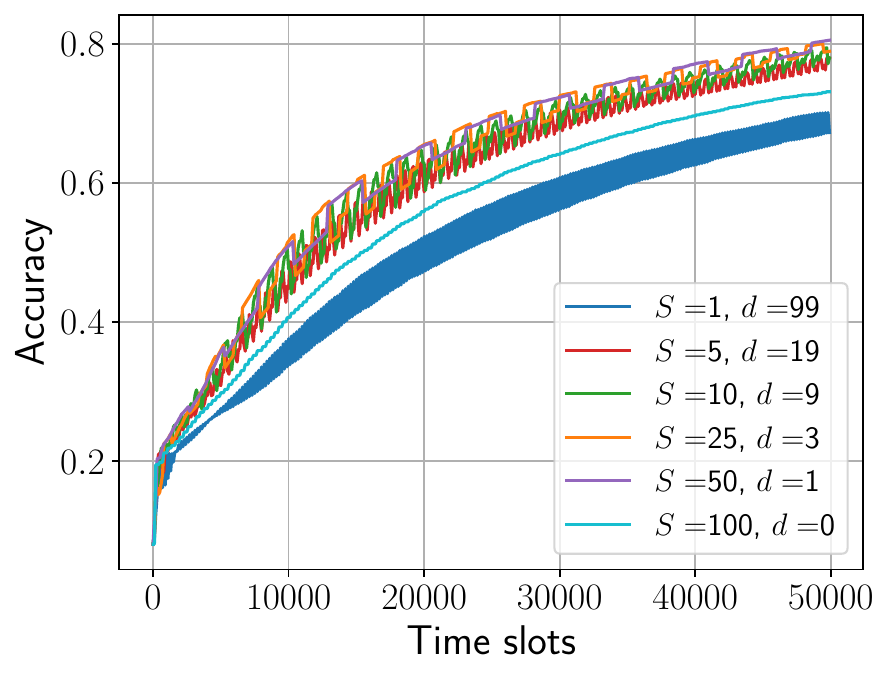}			
		}
		\subfloat[Global loss with respect to training rounds for MNIST dataset]{\label{fig:config_1_0_iterloss}
			\includegraphics[scale=0.35]{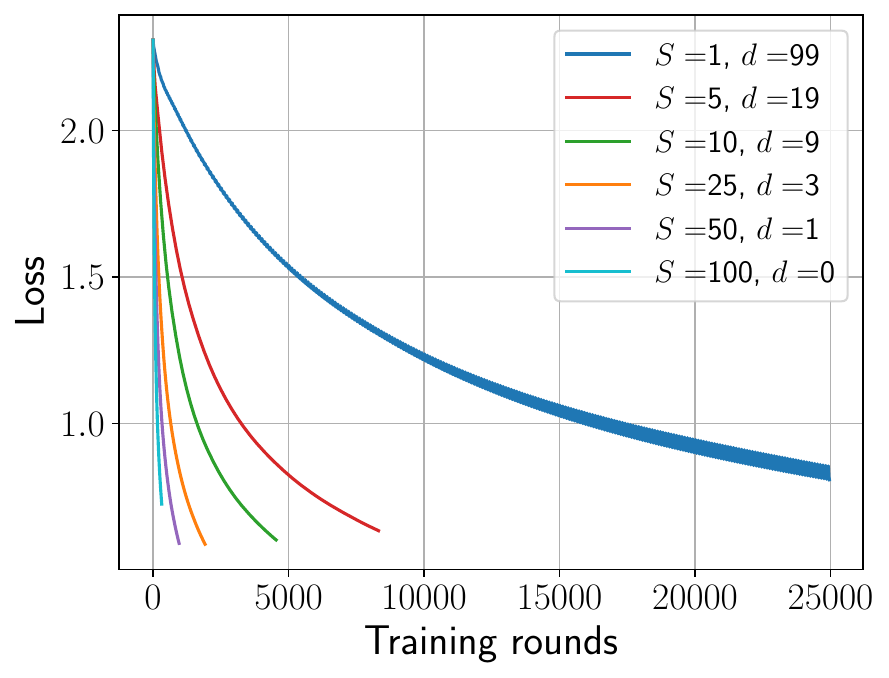}
		}\quad
		\subfloat[Global loss for CIFAR10 dataset]{\label{fig:config_17_loss}
			\includegraphics[scale=0.34]{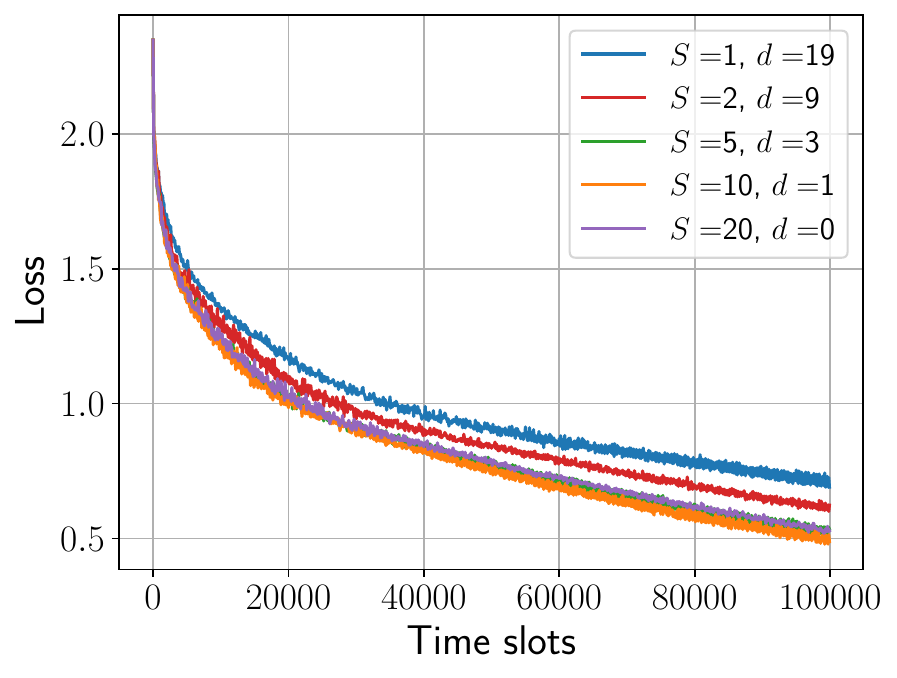}
		}\quad
		\subfloat[Test accuracy for CIFAR10 dataset]{\label{fig:config_17_acc}
			\includegraphics[scale=0.34]{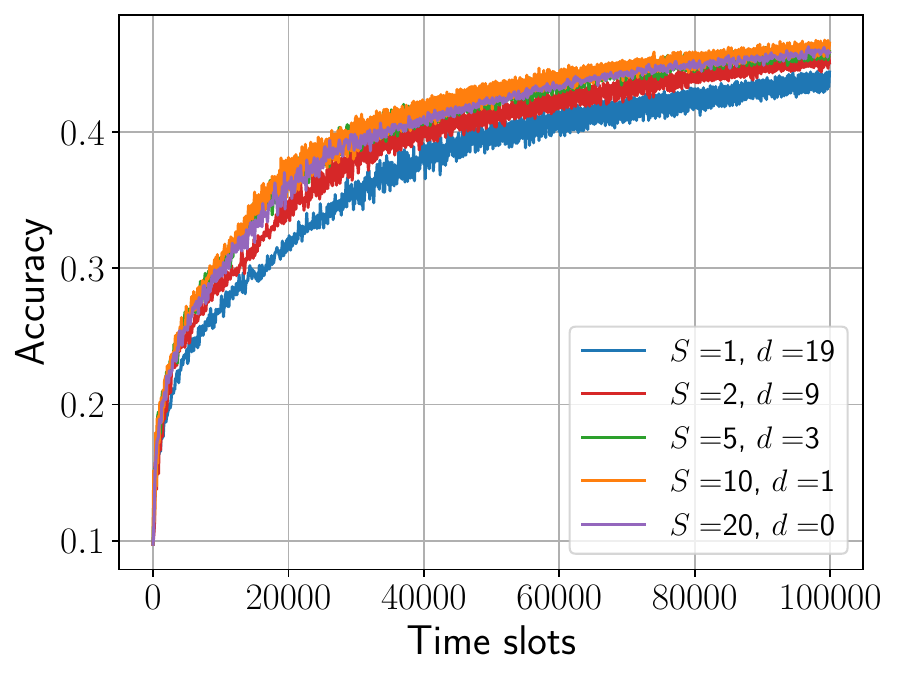}
		}
		\subfloat[Global loss with respect to training rounds for CIFAR10 dataset]{\label{fig:config_17_iterloss}
			\includegraphics[scale=0.34]{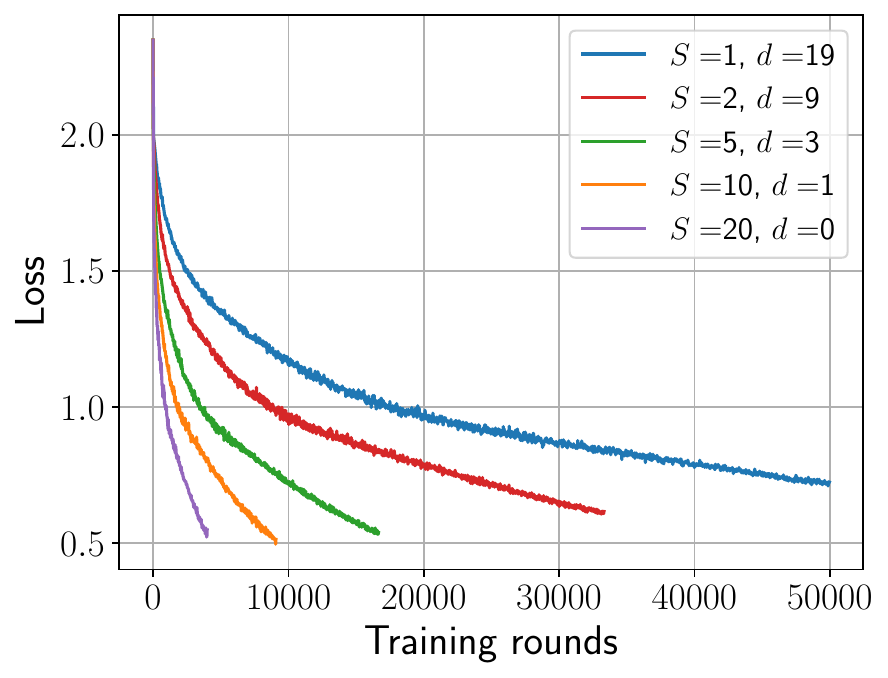}
		}
		\caption{Performance of asynchronous FL over TDMA channel}\label{fig:nonID}
	\end{figure*}

	In this section, we evaluate the performance of asynchronous FL over TDMA channel and IDFL in various environment. We 
	conduct experiment using two datasets: MNIST \cite{Deng2012} and CIFAR10 \cite{Krizhevsky09learningmultiple}.  Convolutional neural network 
	with two convolutional layers is used for our model and cross-entropy loss is calculated for training loss. 
	
	For experiments using MNIST dataset, the experimental configurations are given ast $T=50000$, $N=100$, $S \in \{ 1, 5, 10, 25, 50, 100\}$, and 
	$\eta = 0.01$. Moreover, each IoT device in this scenario has $250$ data samples of a single label chosen randomly. (i.e., $D_n = 250$ for any $n \in 
	\mathcal{N}$.) We also set as $L=5$, $B=64$, 
	$q=6.4$, and $r=1$, which results $\taucomp = 50$ and $\taucomm = S+1$ for given $S$. When we training our model to classify CIFAR10 dataset, 
	we changed the setting as $T=100000$, $N=20$ and $S \in \{1,2,5,10,20\}$. Moreover, we let each IoT device perform $2$ epochs, which is similar 
	to $L=8$ for $250$ data samples with $B=64$. With $q=128$, $\taucomp = 4$ in this scenario.
	
	\subsection{Impact of Delay}
	
	\begin{table}[t!]
		\centering
		\begin{tabular}{|c|c|c|c|c|c|c|}
			\hline
			$S$ & 1  & 5 & 10 & 25 & 50 & 100 \\ \hline
			MNIST & 24976 & 8326 & 4541 & 1922 & 980 & 332 \\ \hline		 
		\end{tabular}\\	
		\vspace{5pt}
		\begin{tabular}{|c|c|c|c|c|c|}
			\hline
			$S$ & 1  & 2 & 5 & 10 & 20 \\ \hline
			CIFAR-10 & 49999 & 33333 & 16667 & 9091 & 4001 \\ \hline
		\end{tabular}
		\vspace{1em}
		\caption{Total number of training rounds}\label{tbl:num_globaliter}      
	\end{table}
	
	We first investigate the impact of TDMA group size $S$ which determines the  staleness of local updates.
	Fig. \ref{fig:nonID} shows the results of Algorithm \ref{alg:async}.  
	Fig. \ref{fig:nonID}-\subref{fig:config_1_0_loss} and Fig. \ref{fig:nonID}-\subref{fig:config_1_0_acc} represent global loss and test accuracy for MNIST 
	dataset. The results for CIFAR10  dataset are shown in Fig. \ref{fig:nonID}-\subref{fig:config_17_loss} and Fig. 
	\ref{fig:nonID}-\subref{fig:config_17_acc}. 
	In the legend of each figure, we represent the TDMA group size $S$ and corresponding staleness of local updates $d = G-1$. 
	
	In Fig. \ref{fig:nonID}-\subref{fig:config_1_0_loss}, the global loss is reduced most rapidly with $S=50$ which corresponds to $d=1$. However, the 
	global loss for 
	$S=100$, which implies global model is updated without delay, is relatively high 
	compared with other $S$. This is because for large $S$, the time slots for single training round becomes large as communication time increases. 
	Therefore, the total number of training round is reduced as $S$ increases given a fixed training time. The total number of global training rounds for different $S$ is shown in Table \ref{tbl:num_globaliter}.
	
	As shown in Fig. \ref{fig:nonID}-\subref{fig:config_17_loss} and Fig. \ref{fig:nonID}-\subref{fig:config_17_acc}, similar trend is observed in the result 
	of CIFAR10 dataset. Due to tradeoff between number of global training rounds and delay of stochastic gradient, $S=5$ shows the minimum loss for 
	CIFAR10 dataset.
	
	If we evaluate the loss of global model with respect to training rounds, it is clearly seen that large $S$ shows rapid 
	decreasing but the number of training rounds allowed given time window for large $S$ is small as shown in Fig. 
	\ref{fig:nonID}-\subref{fig:config_1_0_iterloss} and Fig. \ref{fig:nonID}-\subref{fig:config_17_iterloss} for MNIST and CIFAR-10 dataset, respectively.
	
	Furthermore, it is worth noting that as $S$ increases, the global loss and test accuracy is less fluctuated. 
	As delay becomes smaller, the deviation between local models decreases, which leads to stable reduction of global loss. 
	In the perspective of stability, large $S$ can be preferred for applications which necessitate low fluctuation of global loss.
	
	\subsection{Performance of Intentional Delay Federated Learning}
	In this subsection, we demonstrate the performance of IDFL and compare it with asynchronous FL where the intentional delay is 
	not applied.
	
	\begin{figure*}[t!]
		\centering
		\subfloat[Global loss for MNIST dataset]{\label{fig:config_1_1_loss}
			\includegraphics[scale=0.35]{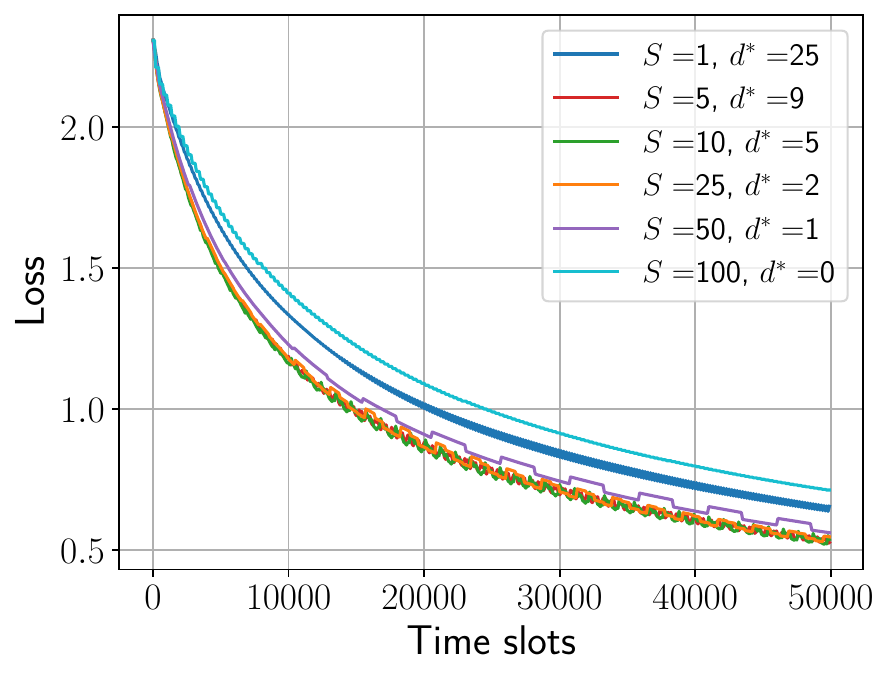}			
		}
		\subfloat[Test accuracy for MNIST dataset]{\label{fig:config_1_1_acc}
			\includegraphics[scale=0.35]{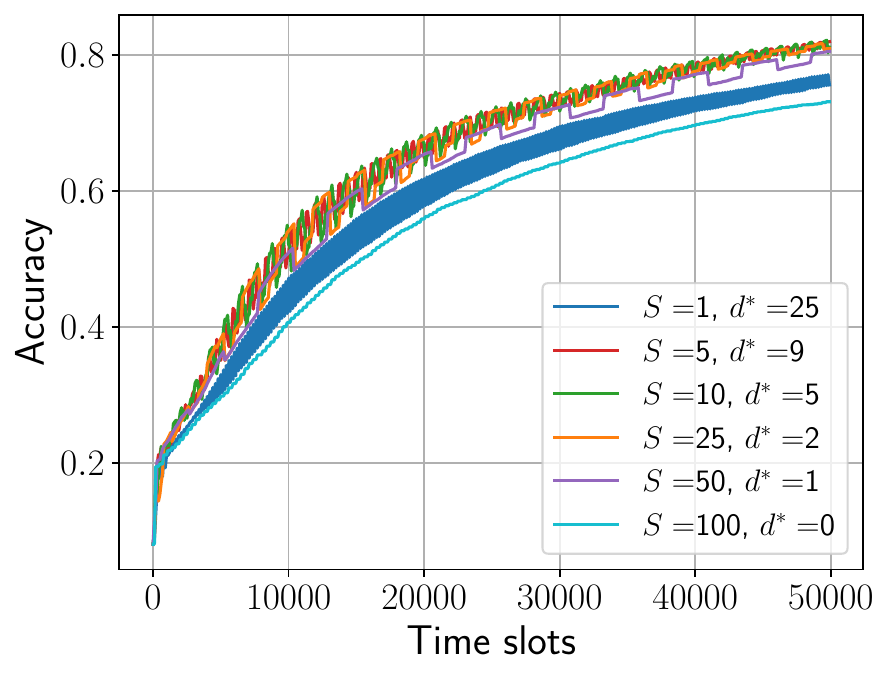}			
		}
		\subfloat[Global loss with respect to training rounds for MNIST dataset]{\label{fig:config_1_1_iterloss}
			\includegraphics[scale=0.35]{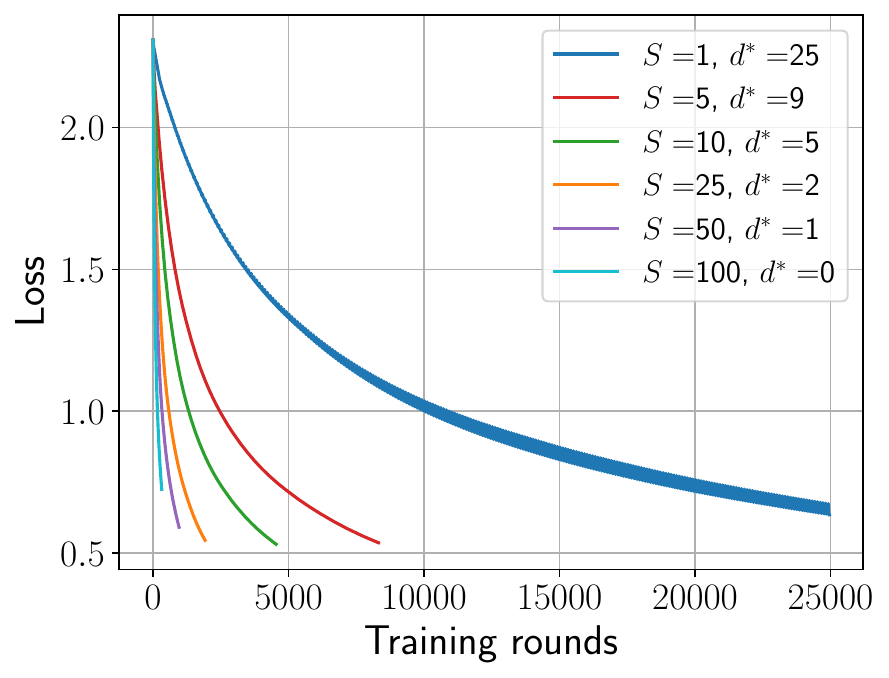}
		}\quad
		\subfloat[Global loss for CIFAR10 dataset]{\label{fig:config_18_loss}
			\includegraphics[scale=0.34]{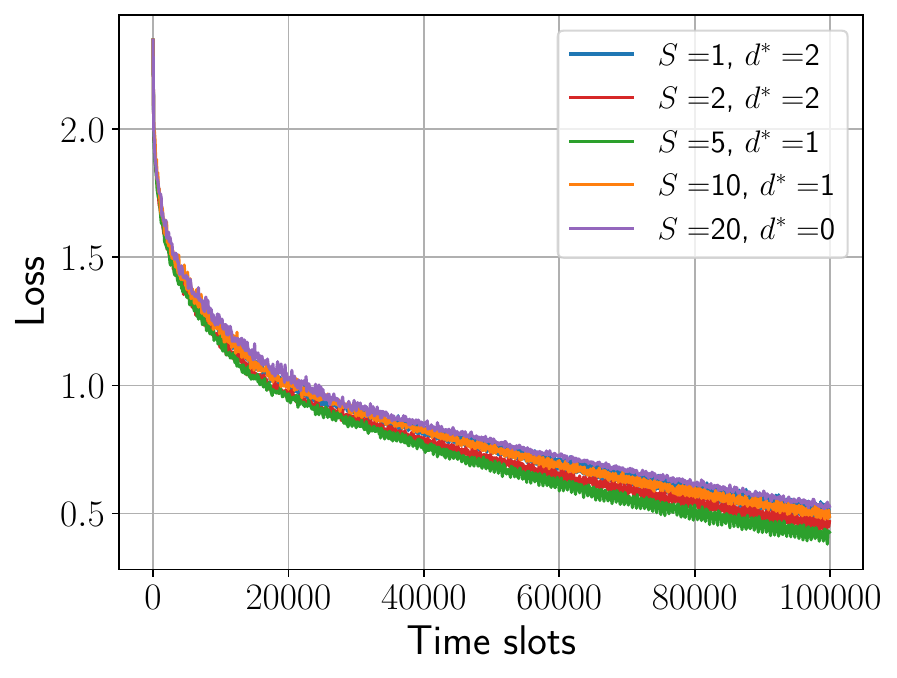}
		}\quad
		\subfloat[Test accuracy for CIFAR10 dataset]{\label{fig:config_18_acc}
			\includegraphics[scale=0.34]{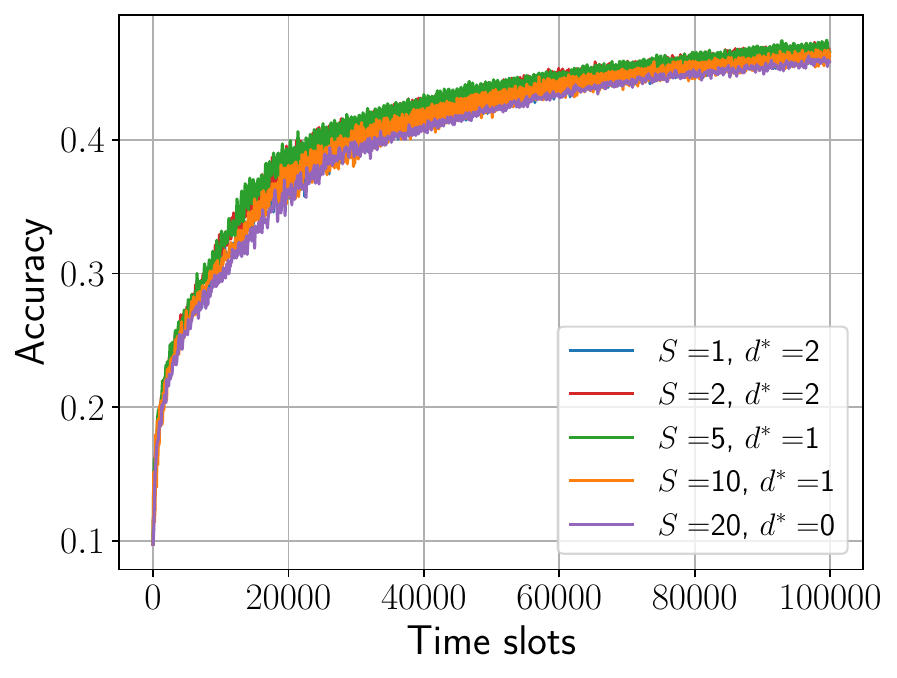}
		}
		\subfloat[Global loss with respect to training rounds for CIFAR10 dataset]{\label{fig:config_18_iterloss}
			\includegraphics[scale=0.34]{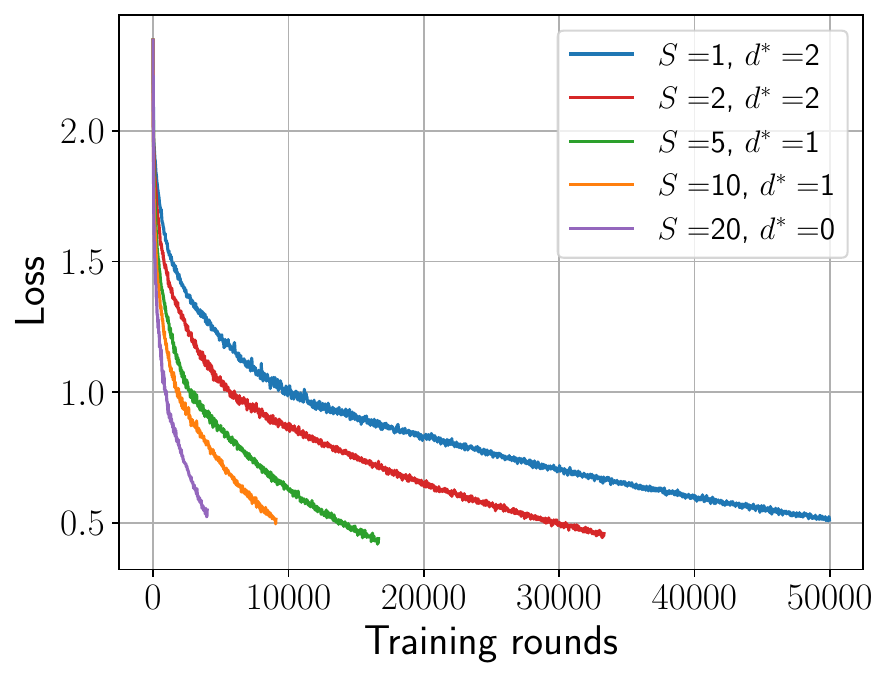}
		}
		\caption{Performance of IDFL}\label{fig:IDFL}
	\end{figure*}
	
	Fig. \ref{fig:IDFL}-\subref{fig:config_1_1_loss} and Fig. \ref{fig:IDFL}-\subref{fig:config_1_1_acc} show global loss and test accuracy of IDFL for 
	MNIST dataset. 
	Compared to Fig. \ref{fig:nonID}-\subref{fig:config_1_0_loss}, IDFL is shown to achieve relatively low global loss due to the reduced outdatedness of local updates.
	For $S=N$, all local updates are computed from the up-to-date model; thus, 
	the intentional delay cannot be applied. Therefore, performance of IDFL and asynchronous FL with $S=N$ is 
	identical. 
	In this context, it is remarkable that the global loss with $S=1$ becomes lower than that with $S=100$ after applying intentional delay.
	
	Although $S=50$ achieves the lowest global loss in Fig. \ref{fig:nonID}-\subref{fig:config_1_0_iterloss}, the minimum global loss is          achieved with $S=10$ when intentional delay is applied. 
	Thanks to the reduced staleness provided by the introduction of intentional delay, we can use a relatively smaller TDMA group size to increase the frequency of global model updates within a given time frame.
	Similarly, for the CIFAR10 dataset, the optimal $S$ changes from $S=10$ to $S=5$ due to the intentional delay, as shown in Fig. 
	\ref{fig:IDFL}-\subref{fig:config_18_iterloss}.
	
	\begin{figure}[t!]
		\centering
		\subfloat[MNIST dataset]{\label{fig:loss_comp_1}
			\includegraphics[scale=0.5]{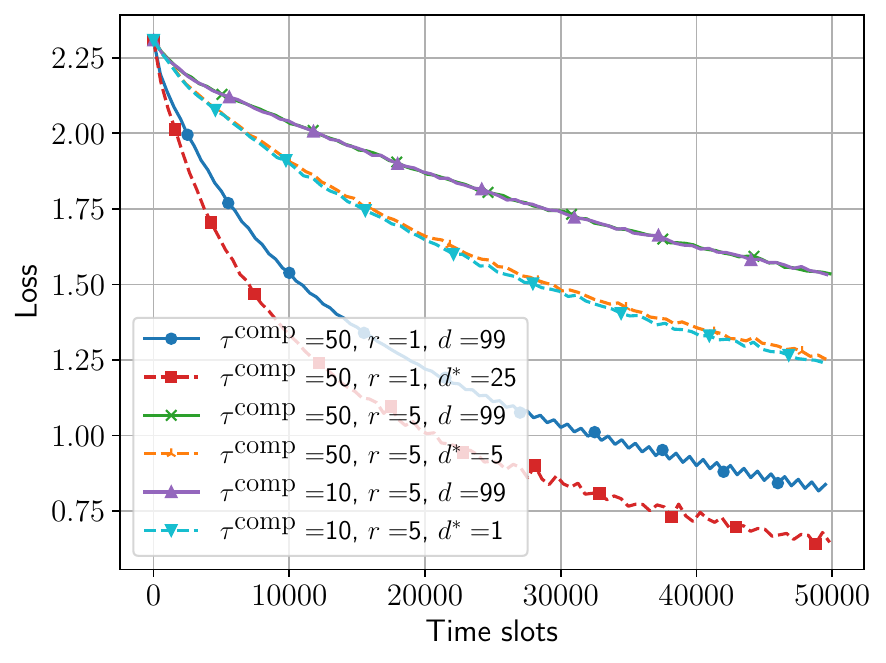}			
		}\quad
		\subfloat[CIFAR10 dataset]{\label{fig:loss_comp_2}
			\includegraphics[scale=0.5]{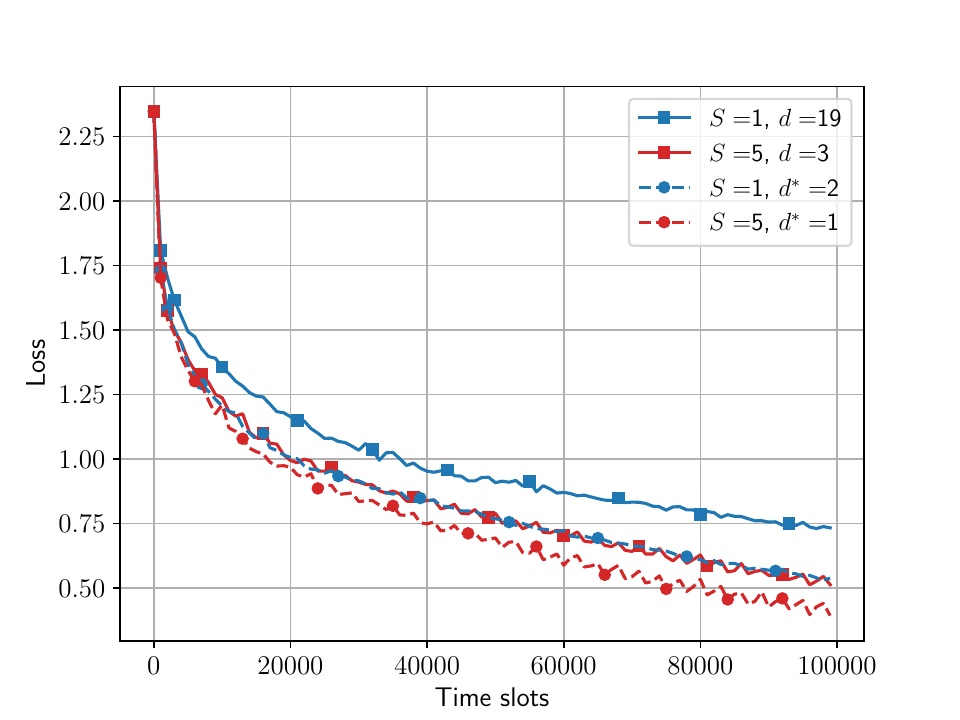}			
		}
		\caption{Comparison of global loss between Asynchronous FL with delayed gradient and IDFL}\label{fig:comp}
	\end{figure}
	
	Moreover, we demonstrate the improvement caused by intentional delay in Fig. \ref{fig:comp}. 
	To enhance the clarity of graph, we plot the markers at interval of 1,000 time slots.
	The amount of reduction is large for small $S$ as the small size of communication group causes larger staleness of local updates.
	In the same vein, for large $S$, small intentional delay is allowed, which leads to low enhancement with the intentional delay.
	
	\subsection{Tradeoff between Computation and Communication Factors}
	
	\begin{figure}[t!]
		\centering
		\includegraphics[scale=0.5]{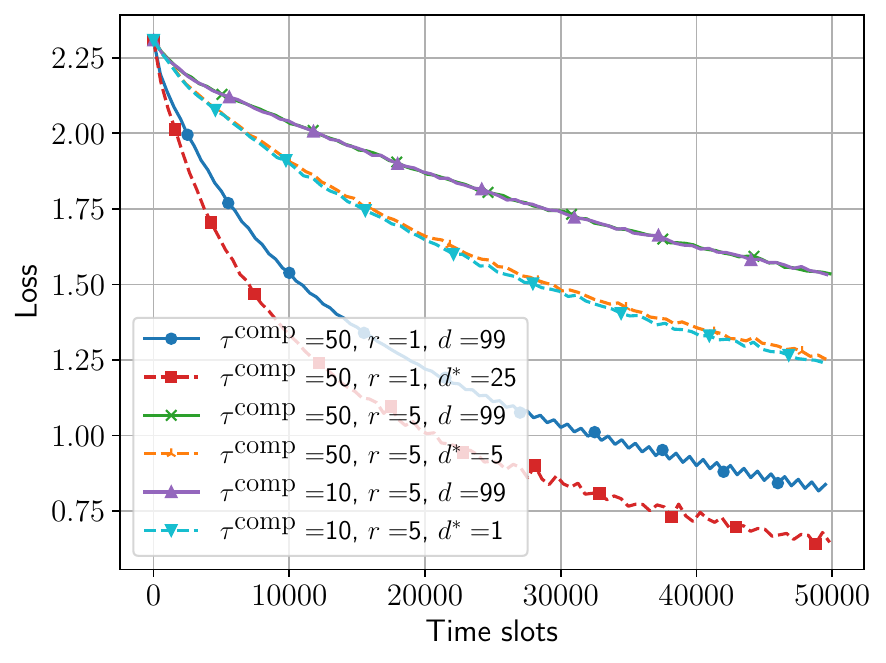}
		\caption{Comparison of global loss between Asynchronous FL with delayed gradient and IDFL with different $\taucomp$ and 
			$r$}\label{fig:tradeoff_1}
	\end{figure}
	
	\begin{figure}[t!]
		\centering
		\includegraphics[scale=0.5]{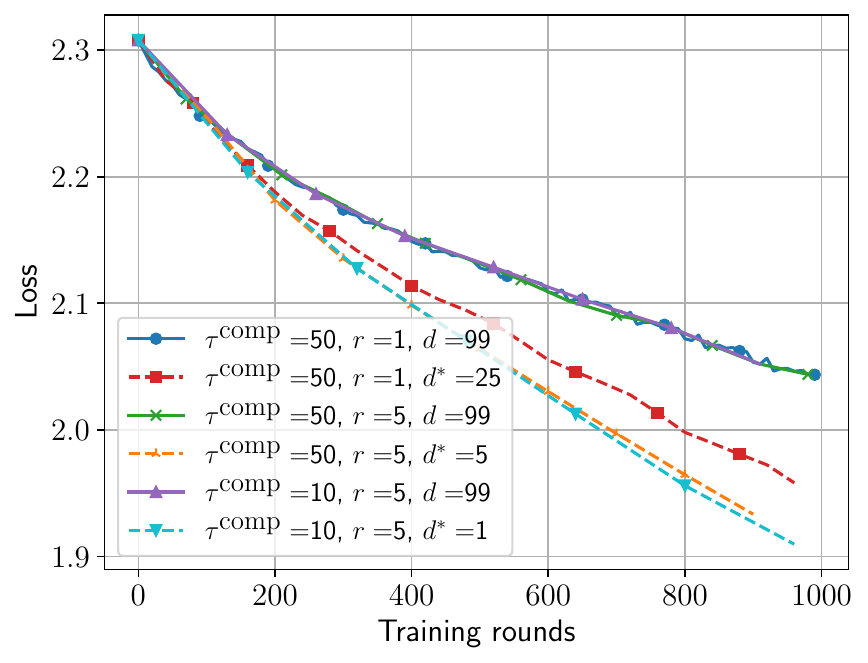}
		\caption{Comparison of global loss between Asynchronous FL with delayed gradient and IDFL with different 
			$\taucomp$ and 
			$r$}\label{fig:tradeoff_2}
	\end{figure}
	
	In Fig. \ref{fig:tradeoff_1} and Fig. \ref{fig:tradeoff_2}, provided that $S=1$, we compare the global loss for different 
	$\taucomp$ and $r$ with respect to time slots and iterations, respectively, training CNN for MNIST dataset.
	Consequently, for $S=1$, when $\frac{\taucomp}{r} = 50$, the maximum intentional delay is obtained as $\alpha = 74$, which 
	leads to $d^* = 25$. Similarly, for $\frac{\taucomp}{r} = 10$ and $\frac{\taucomp}{r} = 2$, $d^* = 5$ and $d^*=1$ 
	are obtained, respectively.
	
	As $r$ increases, the total number of training rounds allowed for given time slots decreases. 
	Hence, global loss for $r=5$ is relatively high compared with that of $r=1$. 
	However, as large $r$ allows larger intentional delay, which results in low effective delay, global loss of $r=5$ with respect to training round can be lower than that of $r=1$.
	
	From \eqref{eq:max_delay}, for a given $S$, the intentional delay $\alpha$ decreases as $\frac{\taucomp}{r}$ 
	increases. Intuitively, large $\frac{\taucomp}{r}$ implies that computation time is relatively longer than 
	communication time. Thus, in order to finish computation to transmit on time for next turn, each IoT device needs to 
	receive updated global model in early iteration. Hence, intentional delay should be small. On the other hand, when
	$\frac{\taucomp}{r}$ is small, IoT devices can finish local computation even if receiving global model in the previous 
	training round of the transmitting iteration. 
	
	\section{Conclusion}\label{sec:conclusion}
	In this paper, we have investigated asynchronous FL in which the local updates of IoT devices are collected using a TDMA scheme.
	We have formulated that the staleness of local updates is determined by the TDMA group size and discovered the trade-off between the staleness of local updates and the frequency of global model updates in asynchronous FL over a TDMA channel.
	Moreover, through convergence analysis, the impact of the outdated local updates on the convergence have been derived. 
	In order to reduce performance degradation from the outdated local updates, we have refined the FL strategy by introducing an intentional delay defined as the number of training rounds to wait for receiving more recent global model. 
	Since the value of intentional delay is dependent on the size of TDMA group, the optimal group size is changed after intentional delay is applied for asynchronous FL over TDMA channel. In general, when the intentional delay is applied, lower size of TDMA group achieves the minimum global loss.
	
	\appendices
	\setcounter{equation}{0}
	\renewcommand{\theequation}{A.\arabic{equation}}
	
	\section{Proof of Lemma \ref{lem:descent_1}} \label{App:lemma1}
	From the $L$-smoothness of global function, we have
	\begin{align}
		f(\mathbf{w}_{k+1}) \leq f(\mathbf{w}^{k}) - \eta \nabla f(\mathbf{w}_k)^\intercal \bar{\mathbf{g}}_k + \frac{L 
			\eta^2}{2} \Vert \bar{\mathbf{g}}_k \Vert^2 
		. \label{eq:conv1}
	\end{align}
	After taking expectation with respect to batch selected at the training round $k$ given previous batches, we can rewrite 
	\eqref{eq:conv1} as
	\begin{align}
		\mathbb{E} \left[ f(\mathbf{w}_{k+1} ) \right] \leq  f(\mathbf{w}_k)  - \eta \nabla f (\mathbf{w}_k)^\intercal \mathbb{E} 
		[ \bar{\mathbf{g}}_k ] + 
		\frac{L \eta^2}{2} \mathbb{E} [ \Vert \bar{\mathbf{g}}_k \Vert^2 ] \label{ineq:conv1}.
	\end{align}
	Moreover, $\nabla f (\mathbf{w}_k)^\intercal \mathbb{E} [ \bar{\mathbf{g}}_k ]$ can be expanded as 
	\begin{align}
		&\nabla f (\mathbf{w}_k)^\intercal \mathbb{E} [ \bar{\mathbf{g}}_k ] = \frac{1}{S}\sum_{n \in 
			\mathcal{S}_k}  \nabla 
		f(\mathbf{w}_k)^\intercal \nabla f_n(\mathbf{w}_{\hat{k}_n}), \\			
		\nonumber & = \frac{1}{2} \Vert \nabla f(\mathbf{w}_k) \Vert^2 + \frac{1}{2S}\sum_{n \in \mathcal{S}_k} 
		\Vert \nabla 
		f_n(\mathbf{w}_{\hat{k}_n}) \Vert^2 \\
		& - \frac{1}{2S}\sum_{n \in \mathcal{S}_k} \Vert \nabla f(\mathbf{w}_k) - \nabla 
		f_n(\mathbf{w}_{\hat{k}_n}) 
		\Vert^2. \label{eq:conv2}
	\end{align}
	Moreover, we can bound $\Vert \nabla f(\mathbf{w}_k) - \nabla f_n(\mathbf{w}_{\hat{k}_n}) \Vert^2$ as
	\begin{align}
		\nonumber \Vert \nabla f(\mathbf{w}_k) - \nabla f_n(\mathbf{w}_{\hat{k}_n}) \Vert^2 \leq & \| \nabla 
		f(\mathbf{w}_k) -  \nabla f (\mathbf{w}_{\hat{k}_n}) \| ^2 \\
		& + \| \nabla f (\mathbf{w}_{\hat{k}_n}) - \nabla f_n(\mathbf{w}_{\hat{k}_n}) \|^2 .
	\end{align}
	Using the assumption \ref{assump_smooth} and \ref{assump_hetero}, we have,
	\begin{align}
		\Vert \nabla f(\mathbf{w}_k) - \nabla f_n(\mathbf{w}_{\hat{k}_n}) \Vert^2 \leq & L^2 \| \mathbf{w}^{k} - 
		\mathbf{w}_{\hat{k}_n} \|^2 + 
		\Gamma^2 . \label{ineq:thm1_temp1}
	\end{align}
	Using \eqref{ineq:thm1_temp1}, \eqref{eq:conv2} can be rewritten as
	\begin{align}
		&\nonumber \nabla f (\mathbf{w}_k)^\intercal \mathbb{E} [ \bar{\mathbf{g}}_k ]  \geq \frac{1}{2} \Vert \nabla 
		f(\mathbf{w}_k) \Vert^2 +  
		\frac{1}{2S}\sum_{n \in \mathcal{S}_k} \Vert \nabla f_n(\mathbf{w}_{\hat{k}_n}) \Vert^2 \\
		& - \frac{\Gamma^2}{2} - \frac{L^2}{2S}\sum_{n \in \mathcal{S}_k} \Vert \mathbf{w}_k - 
		\mathbf{w}_{\hat{k}_n} 
		\Vert^2 .  \label{ineq:conv3}
	\end{align}		
	
	Furthermore, the second moment of stochastic gradients is bounded as
	\begin{align}
		\mathbb{E} [ \Vert \bar{\mathbf{g}}_k \Vert^2 ] & = \mathbb{E} \left[ \left\Vert \frac{1}{S} 
		\sum_{n\in \mathcal{S}_k} \nabla f_n( 
		\mathbf{w}_{\hat{k}_n} ; \mathcal{B}_{n,k}) \right\Vert^2 \right] , \\
		& \leq \frac{1}{S^2} \sum_{n \in \mathcal{S}_k} \mathbb{E} \left[ \Vert \nabla f_n( 
		\mathbf{w}_{\hat{k}_n} ; 
		\mathcal{B}^{\hat{k}_n}_n ) \Vert^2 \right] , \\
		& \leq \frac{\sigma^2}{S B} + \frac{M}{S^2 B} \sum_{n \in \mathcal{S}_k } \Vert \nabla 
		f_n(\mathbf{w}_{\hat{k}_n} ) 
		\Vert^2. \label{ineq:conv4}
	\end{align}
	
	Using \eqref{ineq:conv3} and \eqref{ineq:conv4}, we can further bound \eqref{ineq:conv1} as
	\begin{align}
		\nonumber &\mathbb{E} \left[ f(\mathbf{w}_{k+1} ) \right] \leq  f(\mathbf{w}_k)  - \frac{\eta}{2} \Vert \nabla 
		f(\mathbf{w}_k) \Vert^2 \\
		\nonumber & + \left( \frac{\eta^2 M L}{2 S^2 B }-\frac{\eta}{2S } \right) \sum_{n \in 
			\mathcal{S}_k} \Vert \nabla 
		f_n(\mathbf{w}_{\hat{k}_n}) \Vert^2 + 
		\frac{\eta \Gamma^2}{2} \\
		& + \frac{\eta L^2}{2 S} \sum_{n \in \mathcal{S}_k} \left \| \mathbf{w}_k - 
		\mathbf{w}_{\hat{k}_n} \right \|^2 + \frac{\eta^2 
			\sigma^2  L }{2 SB} .
	\end{align}
	If we take expectation with respect to all batches up to training round $k$, we have
	\begin{align}
		\nonumber &\mathbb{E} \left[ f(\mathbf{w}_{k+1} ) \right] \leq  \mathbb{E}[ f(\mathbf{w}_k)]  - \frac{\eta}{2} 
		\mathbb{E} \left[ \Vert \nabla 
		f(\mathbf{w}_k) \Vert^2 \right] \\
		\nonumber & + \left( \frac{\eta^2 M L}{2 S^2 B}-\frac{\eta}{2S} \right) \sum_{n \in 
			\mathcal{S}_k} \mathbb{E} \left[ 
		\left\| \nabla f_n (\mathbf{w}_{\hat{k}_n} ) \right \|^2 \right] + \frac{\eta \Gamma^2}{2} \\
		& + \frac{\eta L^2}{2 S} \sum_{n \in \mathcal{S}_k} \mathbb{E} \left[ \left \| \mathbf{w}_k - 
		\mathbf{w}_{\hat{k}_n} \right \|^2 \right]  + \frac{\eta^2 	\sigma^2  L }{2 S B}.
	\end{align}

	\setcounter{equation}{0}
	\renewcommand{\theequation}{B.\arabic{equation}}
	
	\section{Proof of Lemma \ref{lem:lemma3}} \label{App:lemma3}
	Since the difference between the two global parameter vectors in different training rounds equals to the sum of the aggregated updates applied from $\hat{k}_n$ to $k-1$, the term can be expanded as
	\begin{align}
		\nonumber &\sum_{n \in }\mathbb{E} \left[ \Vert \mathbf{w}_k - \mathbf{w}_{\hat{k}_n} \Vert^2 \right] \\
		& = \sum_{n \in \mathcal{S}_k}\mathbb{E} \left[ \left\Vert \sum_{j= \hat{k}_n}^{k-1} \frac{\eta}{S} \sum_{n' \in 
			\mathcal{S}_j} \nabla f_{n'}(\mathbf{w}_{\hat{j}_{n'}}  ; \mathcal{B}_{n', \hat{j}_n}) \right\Vert^2 \right] , \nonumber\\
		& \leq \left(\frac{\eta}{S}\right)^2 \sum_{n \in  \mathcal{S}_k} \sum_{j= \hat{k}_n}^{k-1} \sum_{n' \in 
			\mathcal{S}_j} \mathbb{E} \left[ \Vert \nabla f_{n'}(\mathbf{w}_{\hat{j}_{n'}}  ; \mathcal{B}_{n', \hat{j}_{n'}}) \Vert^2 
		\right] , \nonumber\\
		& \nonumber \leq \left(\frac{\eta}{S}\right)^2 \sum_{n \in } \sum_{j= \hat{k}_n}^{k-1} \sum_{n' \in 
			\mathcal{S}_j} \frac{\sigma^2}{B} \\
		& ~~~+ \left(\frac{\eta}{S}\right)^2 \sum_{n \in  \mathcal{S}_k } \sum_{j= \hat{k}_n}^{k-1} \sum_{n' \in 
			\mathcal{S}_j}  \frac{ M}{B}  \mathbb{E} \left[ \Vert \nabla f_{n'}(\mathbf{w}_{\hat{j}_{n'}} ) \Vert^2 \right]  \label{ineq:conv5}
	\end{align}
	Exploiting \eqref{eq:delay} and Lemma \ref{lem:OutdatedW}, we can obtain the following bound.
	\begin{align}
		&\sum_{n \in \mathcal{S}_k }\!\mathbb{E}\!\left[ \Vert \mathbf{w}_k \!-\! \mathbf{w}_{\hat{k}_n} \Vert^2 \right] \nonumber\\
		&\leq\!  \frac{\eta^2 \sigma^2}{S B } \!\!\sum_{n \in \mathcal{S}_k } ( k \!-\! \hat{k}_n) + \frac{\eta^2 M}{S^2 B }\!\! \sum_{n \in \mathcal{S}_k}\! \sum_{j= \hat{k}_n}^{k-1}  \Gamma^2 S \nonumber\\
		&~~~ +\frac{\eta^2 M}{S^2 B } \sum_{n \in \mathcal{S}_k}\!\! \sum_{j= \hat{k}_n}^{k-1} \!\! \sum_{n' \in 
			\mathcal{S}_j} \mathbb{E} \left[  \Vert \nabla f (\mathbf{w}_{\hat{j}_{n'}} ) 
		\Vert^2 \right] \nonumber\\
		& =  \frac{\eta^2 \left(\sigma^2 + M \Gamma^2 \right) }{S B } \sum_{n \in \mathcal{S}_k } d_{n,k} \nonumber \\
		&~~~+\frac{\eta^2 M}{S^2 B } \sum_{n \in \mathcal{S}_k} \!\! \sum_{j= \hat{k}_n}^{k-1} \!\! \sum_{n' \in 
			\mathcal{S}_j} \mathbb{E}\! \left[  \Vert \nabla f (\mathbf{w}_{\hat{j}_{n'}} ) 
		\Vert^2 \right]. \nonumber
	\end{align}

	\setcounter{equation}{0}
	\renewcommand{\theequation}{C.\arabic{equation}}
	
	\section{Proof of Lemma \ref{lem:sq_norm_dgrad}} \label{App:lemma4}
	
	First, let us consider $\sum_{k=0}^{K} \sum_{n \in \mathcal{S}_k} \mathbb{E} \left[ \Vert \nabla f(\mathbf{w}_{\hat{k}_n}) \Vert^2 
	\right]$. Note that the gradients transmitted in training round $k \leq G-1$ are generated with respect to initial global 
	model and the gradients transmitted in training round $k \geq G$ are generated with respect to model of training round $k- G +1$. 
	Hence,  we can rewrite 
	$\sum_{k=0}^{K} \sum_{n \in \mathcal{S}_k} \mathbb{E} \left[ \Vert \nabla f(\mathbf{w}_{\hat{k}_n}) \Vert^2 \right]$ for $K \geq G$ as  
	\begin{align}
		&\sum_{k=0}^{K} \sum_{n \in \mathcal{S}_k} \mathbb{E} \left[ \Vert \nabla f(\mathbf{w}_{\hat{k}_n}) \Vert^2 \right] \nonumber \\
		&= N \mathbb{E} 
		\left[ 
		\Vert \nabla f (\mathbf{w}_0) \Vert ^2 \right] + \sum_{k = 1}^{K-G+1} S
		\mathbb{E} \left[ \Vert \nabla f ( \mathbf{w}^{k}) \Vert^2 \right] 
		\label{eq:conv1_7}
	\end{align}  
	Furthermore, by adding additional terms, we can bound \eqref{eq:conv1_7} as
	\begin{align}
		\nonumber \sum_{k=0}^{K} \sum_{n \in \mathcal{S}_k} &\mathbb{E} \left[ \Vert \nabla f(\mathbf{w}_{\hat{k}_n}) \Vert^2 \right] \leq (N - 
		S)  \Vert \nabla f(\mathbf{w}_0) \Vert^2  \\
		& + S  \sum_{k=0}^K \mathbb{E} \left[ \Vert \nabla f(\mathbf{w}_k) \Vert^2 \right] \label{eq:conv1_11}.
	\end{align}
	
	Now, we consider $\sum_{k=0}^K \sum\limits_{n \in\mathcal{S}_k} \sum_{j = \hat{k}_n}^{k-1} \sum\limits_{n' \in 
		\mathcal{S}_j} \mathbb{E} 
	\left[ \Vert \nabla f (\mathbf{w}_{\hat{j}_{n'}} ) \Vert ^2 \right] $.  From the definition of $\hat{k}_n$ and $\hat{j}_n$, we have
	\begin{align}
		&\sum_{k=0}^K \sum_{n \in\mathcal{S}_k} \sum_{j = \hat{k}_n}^{k-1} \sum\limits_{n' \in \mathcal{S}_j} \mathbb{E} 
		\left[ \Vert \nabla f (\mathbf{w}_{\hat{j}_{n'}} ) \Vert ^2 \right] \nonumber \\
		& = \sum_{k=0}^K \sum_{n \in\mathcal{S}_k} \sum_{j = k - d_{n,k}}^{k-1} \sum\limits_{n' \in \mathcal{S}_j} 
		\mathbb{E} 
		\left[ \Vert \nabla f 	(\mathbf{w}_{j - d^j_{n'}}) \Vert ^2 \right] . \label{eq:lastterm}
	\end{align}
	From \eqref{eq:delay}, we have $k - d_{n,k} = (k - G +1)^+$ and $j - d_{n'}^j = (j - G+1)^+$ for any $n \in \mathcal{S}_k$ and 
	$n' \in \mathcal{S}_j$ where $(x)^+ $ is rectified linear unit defined as 
	$(x)^+= \max \{ x, 0\}$. Thus, we can rewrite \eqref{eq:lastterm} as
	\begin{align}
		\nonumber &\sum_{k=0}^K \sum_{n \in\mathcal{S}_k} \sum_{j = k - d_{n,k}}^{k-1} \sum\limits_{n' \in \mathcal{S}_j} 
		\mathbb{E} 
		\left[ \Vert \nabla f 	(\mathbf{w}_{j - d^j_{n'}}) \Vert ^2 \right]\\
		&= S^2 \sum_{k=0}^K  \sum_{j=(k-G+1)^+}^{k-1} \mathbb{E} \left[  \| \nabla f( \mathbf{w}_{(j- G +1)^+}) \|^2 \right] .
	\end{align} 
	
	As $j \leq k-1$, $j$ is less than $G-1$ if $k \leq G-1$, Hence, we have $(k-G+1)^+ = (j-G+1)^+ = 0$ for $k \leq G-1$. Consequently,
	\begin{align}
		\nonumber & \sum_{k=0}^{d-1} \sum_{j = (k-G+1)^+}^{k-1} \mathbb{E} \left[ \| \nabla 
		f(\mathbf{w}_{(j-G+1)^+}) \|^2 \right] \\
		&=\sum_{k=0}^{G-1} \sum_{j = 
			0}^{k-1} \mathbb{E} \left[ \| \nabla f(\mathbf{w}_0) \|^2 \right] , \\
		& = \frac{G (G-1)}{2} \mathbb{E} \left[ \| \nabla f(\mathbf{w}_0) \|^2 \right] . \label{eq:last_1}
	\end{align}
	
	When $G \leq k \leq 2G-3$, we have $ (k-G+1)^+ = k-G+1$ but $(j-G+1)^+$ can still be $0$ depending on the range of $j$. In this case, we can 
	derive as follows.
	\begin{align}
		&\sum_{k=d}^{2G-3}  \sum_{j = (k-G+1)^+}^{k-1} \mathbb{E} \left[ \| \nabla f(\mathbf{w}_{(j-d+1)^+}) \|^2 \right] \nonumber \\ 
		& = \sum_{k=d}^{2G-3} \sum_{j = k-G+1}^{k-1} \mathbb{E} \left[ \| \nabla f(\mathbf{w}_{(j-d+1)^+}) \|^2 \right] \\
		& = \sum_{k=d}^{2G-3}  \left[ \sum_{j = k-G+1}^{G-1} \mathbb{E} \left[ \| \nabla f(\mathbf{w}_0) \|^2 \right] + \sum_{j = G}^{G-1} \mathbb{E} \left[ 
		\| 
		\nabla f(\mathbf{w}^{j-G+1}) \|^2 \right] \right] , \\
		& = \sum_{k=d}^{2G-3}  \left[ (G-1) \mathbb{E} \left[ \| \nabla f(\mathbf{w}_0) \|^2 \right] + \sum_{j = 1}^{k-G} \mathbb{E} \left[ \| 
		\nabla f(\mathbf{w}_j) \|^2 \right] \right] ,\\
		& = (G-1)(G-2)  \mathbb{E} \left[ \| \nabla f(\mathbf{w}_0) \|^2 \right] +\sum_{k=G+1}^{2G-3}   \sum_{j = 1}^{k-G} \mathbb{E} \left[ \| 
		\nabla f(\mathbf{w}_j) \|^2 \right] 
	\end{align}
	
	In addition to that,  by adding $\sum_{k=d+1}^{2G-3} \sum_{j=k-G+1}^{G-3} \mathbb{E} \left[ \| \nabla f(\mathbf{w}_j) \|^2 \right]$, which is strictly 
	positive, we have 
	\begin{align}
		&\sum_{k=d}^{2G-3}  \sum_{j = (k-G+1)^+}^{k-1} \mathbb{E} \left[ \| \nabla f(\mathbf{w}_{(j-G+1)^+}) \|^2 \right] \nonumber \\ 
		& \leq (G-1)(G-2)  \mathbb{E} \left[ \| \nabla f(\mathbf{w}_0) \|^2 \right] + \sum_{k=G+1}^{2G-3}   \sum_{j = 1}^{G-3} \mathbb{E} \left[ \| 
		\nabla f(\mathbf{w}_j) \|^2 \right] , \\
		& = (G-1)(G-2)  \mathbb{E} \left[ \| \nabla f(\mathbf{w}_0) \|^2 \right] + (G-3)   \sum_{j= 1}^{G-3} \mathbb{E} \left[ \| 
		\nabla f(\mathbf{w}_j) \|^2 \right]  \label{eq:last_2}
	\end{align}

	For $k \geq 2G - 2$, we have $j \geq G-1$. Thus, $(k - G + 1)^+ = k - d +1$ and $(j -G +1)^+ = j - G +1$. Therefore,
	\begin{align}
		\nonumber &\sum_{k=0}^K \sum_{n \in\mathcal{S}_k} \sum_{j = \hat{k}_n}^{k-1} \sum\limits_{n' \in \mathcal{S}_j} 
		\mathbb{E} 
		\left[ \Vert \nabla f 
		(\mathbf{w}_{\hat{j}_{n'}} ) \Vert ^2 \right] \\
		& = S^2 \sum_{k=0}^{2G-3} \sum_{j=(k-G+1)^+}^{k-1} \mathbb{E} \left[  \| \nabla f( \mathbf{w}_{(j -G +1)^+}) \|^2 \right]  \nonumber \\
		& ~~~+ S^2 \sum_{k=2G-2}^{K} \sum_{j=k-G+1}^{k-1} \mathbb{E} \left[  \| \nabla f( \mathbf{w}_{j - G + 1}) \|^2 \right] 
	\end{align}
	Moreover, for each $k$  if we add $\sum_{j=0}^{k -G +1} \mathbb{E} \left[  \| \nabla f( \mathbf{w}_{j - G +1}) \|^2 \right] $ and $ \sum_{j=k}^{K-G+1} 
	\mathbb{E} \left[  \| \nabla f( \mathbf{w}_{j -G +1}) \|^2 \right] $ to $\sum_{k=2G-2}^{K} \sum_{j=k-G+1}^{k-1} \mathbb{E} \left[  \| \nabla f( 
	\mathbf{w}_{j -G +1}) \|^2 \right] $, we can obtain the following bound.
	\begin{align}
		\sum_{k=2G-2}^{K} \sum_{j=k-G+1}^{k-1} \mathbb{E} \left[  \| \nabla f( \mathbf{w}_{j -G +1}) \|^2 \right]  \leq \sum_{k=0}^{K-G-1} (G-1) 
		\mathbb{E} \left[ \| \nabla f(\mathbf{w}_k) \|^2 \right] . \label{eq:last_3}
	\end{align}
	Combining \eqref{eq:last_1}, \eqref{eq:last_2}, and \eqref{eq:last_3}, the following upper bound is found.
	\begin{align}
		&\sum_{k=0}^K \sum_{n \in\mathcal{S}_k} \sum_{j = \hat{k}_n}^{k-1} \sum\limits_{n' \in \mathcal{S}_j} \mathbb{E} 
		\left[ \Vert \nabla f (\mathbf{w}_{\hat{j}_{n'}} ) \Vert ^2 \right] \nonumber \\
		\leq & S^2 \left(  \frac{G (G-1)}{2} \mathbb{E} \left[ \| \nabla f(\mathbf{w}_0) \|^2 \right] \right. \nonumber \\
		& \left.+ (G-1)(G-2)  \mathbb{E} \left[ \| \nabla f(\mathbf{w}_0) \|^2 \right] + (G-3)   \sum_{j= 1}^{G-3} \mathbb{E} \left[ \| 
		\nabla f(\mathbf{w}_j) \|^2 \right] \right.  \nonumber \\
		& \left. + \sum_{k=0}^{K-G-1} (G-1) 
		\mathbb{E} \left[ \| \nabla f(\mathbf{w}_k) \|^2 \right] . \right) 
	\end{align}
	After some manipulation, we have
	\begin{align}
		&\sum_{k=0}^K \sum_{n \in\mathcal{S}_k} \sum_{j = \hat{k}_n}^{k-1} \sum\limits_{n' \in \mathcal{S}_j} \mathbb{E} 
		\left[ \Vert \nabla f (\mathbf{w}_{\hat{j}_{n'}} ) \Vert ^2 \right] \nonumber \\
		& \leq 2 G S^2 \left( G \left[ \| \nabla f(\mathbf{w}_0)\|^2 \right] + \sum_{k=0}^{K} \mathbb{E} \left[ \| \nabla 
		f(\mathbf{w}_k)\|^2\right] \right) 
	\end{align}

	\setcounter{equation}{0}
	\renewcommand{\theequation}{D.\arabic{equation}}
	
	\section{Proof of Theorem \ref{thm:convergence}} \label{App:theorem1}
	%%%%%%%%%%%%%%%%
	If we apply the upper bound obtained from Lemma \ref{lem:sq_norm_dgrad} to \eqref{ineq:descent_3}, we have
	\begin{align}
		&\frac{\eta}{K+1}  \left(2  - \frac{\eta M L}{ 
			SB} -  \frac{\eta^2 L^2 M G }{S B }  \right) \sum_{k=0}^K \mathbb{E} \left[ \left \| \nabla 
		f(\mathbf{w}_k) \right\|^2 
		\right] \nonumber \\
		& \leq  \frac{\eta^2 ML G ( G + 2G^2 \eta L -1) - \eta N ( \eta G B - 1) }{N B 
			(K+1)} \Vert\nabla f(\mathbf{w}_0) \Vert^2   
		\nonumber \\
		&~~~ +  \frac{2}{K+1} \mathbb{E}\left[ f(\mathbf{w}_0)  -  f(\mathbf{w}_{k+1} )\right] + \frac{\eta^2 L \left( \Gamma^2 M +  \sigma^2 \right) }{ SB}  \nonumber  \\
		&~~~ +  \frac{\eta^3 L^2 
			\left(\sigma^2 + M \Gamma^2 \right) }{S  B (K+1) } \left(  (G-1) \left(  K - \frac{G}{2} +1  
		\right) \right) . \label{ineq:descent_4}
	\end{align}
	Let us define some constants as
	\begin{align}
		Q_1 &\delequal \eta \left(2 - \frac{\eta M L }{ 
			SB} -  \frac{\eta^2 L^2 M N }{S^2 B }  \right) , \nonumber\\
		Q_2 & \delequal \frac{\eta^2  ML G( G + 2G^2 \eta L -1) - \eta N ( \eta G B - 1) }{N B 
			(K+1)} \Vert\nabla f(\mathbf{w}_0) \Vert^2 , \nonumber\\
		Q_3 & \delequal  \frac{2 \mathbb{E}\left[ f(\mathbf{w}_0)  -  f(\mathbf{w}_{k+1} )\right] }{K+1}, \nonumber\\
		\nonumber Q_4 & \delequal  +  \frac{\eta^3 L^2 
			\left(\sigma^2 + M \Gamma^2 \right) }{S^2 B (K+1) } \left( S (G-1) \left(  K - \frac{G}{2} +1  
		\right) \right) \nonumber\\
		&~~~+ \frac{\eta^2 L \left( \Gamma^2 M +  \sigma^2 \right) }{ SB} .\nonumber
	\end{align}
	Note that $\eta \leq \beta$ guarantees that $C_1 >0$. 	Using $Q_1, Q_2, Q_3$ and $Q_4$, we can rewrite \eqref{ineq:descent_4} as
	\begin{align}
		\frac{1}{K+1} \sum_{k=0}^K  \mathbb{E} \left[ \Vert \nabla f (\mathbf{w}_k) \Vert^2 \right] 
		\leq\frac{Q_2}{Q_1} + \frac{Q_3}{Q_1} + \frac{Q_4}{Q_1} \nonumber
	\end{align}	
	Since $\eta\le \beta/\sqrt{K+1}$, each terms can be bounded as, using big $\mathcal{O}$ notation,
	\begin{align}
		\frac{Q_2}{Q_1} &= \mathcal{O}\left( \frac{G^2}{\sqrt{K} }\right),  \nonumber\\
		\frac{Q_3}{Q_1} & = \mathcal{O} \left( \frac{1}{\sqrt{K}} \right),  \nonumber\\
		\frac{Q_4}{Q_1} & = \mathcal{O} \left( \frac{G^2}{ K} + \frac{G}{\sqrt{K}} \right).\nonumber
	\end{align} 
	
	Since the most dominant term is $\mathcal{O}\left( \frac{G^2}{\sqrt{K} }\right)$, we have 
	\begin{align}
		\frac{1}{K+1} \sum_{k=0}^K  \mathbb{E} \left[ \Vert \nabla f (\mathbf{w}_k) \Vert^2 \right]  = \mathcal{O} \left(   
		\frac{G^2}{\sqrt{K}}  \right).
	\end{align}

	% by themselves when using endfloat and the captionsoff option.
	\ifCLASSOPTIONcaptionsoff
	\newpage
	\fi

	% trigger a \newpage just before the given reference
	% number - used to balance the columns on the last page
	% adjust value as needed - may need to be readjusted if
	% the document is modified later
	%\IEEEtriggeratref{8}
	% The "triggered" command can be changed if desired:
	%\IEEEtriggercmd{\enlargethispage{-5in}}
	
	% references section	
	\bibliography{references}
	\bibliographystyle{IEEEtran}
	% can use a bibliography generated by BibTeX as a .bbl file
	% BibTeX documentation can be easily obtained at:
	% http://mirror.ctan.org/biblio/bibtex/contrib/doc/
	% The IEEEtran BibTeX style support page is at:
	% http://www.michaelshell.org/tex/ieeetran/bibtex/
	%\bibliographystyle{IEEEtran}
	% argument is your BibTeX string definitions and bibliography database(s)
	%\bibliography{IEEEabrv,../bib/paper}
	%
	% <OR> manually copy in the resultant .bbl file
	% set second argument of \begin to the number of references
	% (used to reserve space for the reference number labels box)
	%\begin{thebibliography}{1}
	%\end{thebibliography}
	
	% biography section
	% 
	% If you have an EPS/PDF photo (graphicx package needed) extra braces are
	% needed around the contents of the optional argument to biography to prevent
	% the LaTeX parser from getting confused when it sees the complicated
	% \includegraphics command within an optional argument. (You could create
	% your own custom macro containing the \includegraphics command to make things
	% simpler here.)
	%\begin{IEEEbiography}[{\includegraphics[width=1in,height=1.25in,clip,keepaspectratio]{mshell}}]{Michael Shell}
	% or if you just want to reserve a space for a photo:

	%\vfill
	
	% Can be used to pull up biographies so that the bottom of the last one
	% is flush with the other column.
	%\enlargethispage{-5in}

	% that's all folks
\end{document}